\documentclass[12pt]{article}
\usepackage{latexsym,amsmath}
\usepackage[dvips]{graphicx}
\usepackage{lscape}
\usepackage{color}
\usepackage[pagewise]{lineno}

\usepackage{graphicx} 
\usepackage {epstopdf}
\usepackage {subfigure}
\usepackage{hyperref}
\usepackage{booktabs} 
\usepackage{amsmath}
\usepackage{multicol}
\usepackage{multirow}
\usepackage{xcolor}
\usepackage{float}
\usepackage{amsthm}
\usepackage{amsfonts}
\usepackage[margin=1in]{geometry}
\usepackage[toc]{appendix}
\usepackage{stackengine}
\usepackage{bbold}
\RequirePackage[OT1]{fontenc}
\usepackage[authoryear]{natbib}
\usepackage{adjustbox}
\usepackage{amsmath,amssymb,amsthm,bm,latexsym}
\usepackage{graphicx,psfrag,epsf}
\usepackage[ruled,vlined]{algorithm2e}
\usepackage{epigraph,xcolor}
\usepackage{enumerate}
\usepackage{url}
\usepackage{booktabs}
\usepackage{multirow}
\usepackage{hyperref}
\usepackage{pifont}
\hypersetup{colorlinks=true,citecolor=blue,
	pdfpagemode=FullScreen,linktocpage=true}
\usepackage{cleveref}
\usepackage{amsfonts}
\usepackage{amsmath}
\usepackage{orcidlink}
\usepackage{amsthm}
\usepackage[T1]{fontenc}
\usepackage{multirow}
\usepackage{array}
\usepackage{booktabs}

\newtheorem{theorem}{Theorem}
\newtheorem{proposition}{Proposition}

\def\doublespace{\baselineskip=24pt}

\newcommand{\uiota}             {\mbox{\boldmath$\uiota$}}

\def\beq{\begin{equation}}
\def\eeq{\end{equation}}
\def\beqa{\begin{eqnarray}}
\def\eeqa{\end{eqnarray}}
\def\beqan{\begin{eqnarray*}}
\def\eeqan{\end{eqnarray*}}

\def\bc{\begin{center}}
\def\ec{\end{center}}
\def\btable{\begin{table}[htbp]}
\def\etable{\end{table}}
\def\bfig{\begin{figure}[htbp]}
\def\efig{\end{figure}}

\def\bi{\begin{itemize}}
\def\ei{\end{itemize}}
\def\E{\mathbb{E}}

\def\P{\mathbb{P}}

\def\R{\mathbb{R}}

\newcommand{\convD}{\xrightarrow[]{\mathcal{L}}}
\newcommand{\convP}{\xrightarrow[]{\mathcal{P}}}

\renewcommand{\P}{\mathbb{P}}

\newcommand{\RNum}[1]{\uppercase\expandafter{\romannumeral #1\relax}}

\numberwithin{equation}{section}  
\newtheoremstyle{general}
{3mm} 
{3mm} 
{\it} 
{} 
{\bfseries} 
{.} 
{.5em} 
{} 

\theoremstyle{general}
\newtheorem{definition}{Definition}

\begin{document}

\doublespace
\noindent{\huge\bf  Application of Random Matrix Theory in High-Dimensional Statistics
 }\\
 \
 
\noindent Swapnaneel Bhattacharyya,$^{1,2}$, Srijan Chattopadhyay,$^{1,3}$, Sevantee Basu,$^{1,4}$  \\
\
\noindent{$^1$ Indian Statistical Institute, 203 B.T. Road, Kolkata-700108, India}\\
\
\noindent{$^2$
\href{mailto:swapnaneelbhattacharyya@gmail.com}{swapnaneelbhattacharyya@gmail.com}\\
\
\noindent{$^3$ \href{mailto:srijanchatterjee123456789@gmail.com}{srijanchatterjee123456789@gmail.com}}\\
\
\noindent{$^4$ \href{mailto:sevanteebasu@gmail.com}{sevanteebasu@gmail.com}}

\noindent{\bf Abstract}: This review article provides an overview of random matrix theory (RMT) with a focus on its growing impact on the formulation and inference of statistical models and methodologies. Emphasizing applications within high-dimensional statistics, we explore key theoretical results from RMT and their role in addressing challenges associated with high-dimensional data. The discussion highlights how advances in RMT have significantly influenced the development of statistical methods, particularly in areas such as covariance matrix inference, principal component analysis (PCA), signal processing, and changepoint detection, demonstrating the close interplay between theory and practice in modern high-dimensional statistical inference. 


\newpage 

\tableofcontents

\newpage

\section{Introduction}
\label{sec:intro}

In recent years, the field of statistics has seen a significant shift driven by the rapid generation of large, complex datasets across diverse disciplines such as genomics, atmospheric science, communications, biomedical imaging, and economics. These datasets, often high-dimensional due to their representation in standard coordinate systems, pose challenges that extend beyond the scope of classical multivariate statistical methods. This evolving landscape has necessitated the integration of advanced mathematical frameworks, including convex analysis, Differential geometry, topology and combinatorics, into statistical methodologies. Among these, random matrix theory has emerged as a powerful tool for addressing key theoretical and practical problems in the analysis of high-dimensional data.

In this review article, we focus on several application areas of random matrix theory (RMT) in high-dimensional statistics. These include
 problems in dimension reduction, hypothesis testing for high-dimensional data, regression analysis, and covariance estimation. We also briefly describe the important role played by RMT in enabling certain theoretical analyses in wireless communications and changepoint detection. The challenges posed by high-dimensional data have sparked renewed interest in several classical phenomena within random matrix theory (RMT). Among these, the concept of universality holds particular significance, offering insights into the applicability of statistical techniques beyond the traditional framework based on the multivariate Gaussian distribution. This article focuses on aspects of RMT that are most relevant to statistical questions in this context. In particular, attention is directed toward the behavior of the bulk spectrum, represented by the empirical spectral distribution, and the edge of the spectrum, characterized by the extreme eigenvalues of random matrices. Given the central role of the sample covariance matrix in multivariate analysis, a significant portion of this work is devoted to examining its spectral properties and their implications for statistical applications.

A detailed discussion of these topics is present in \citet{bai2010spectral}, \citet{couillet2022random}. Also \citet{johnstone2006high}, \citet{paul2014random} discuss many RMT-based approaches to modern high-dimensional problems. Though RMT has a wide variation of applications beyond statistics, e.g. wireless communications, finance, and econometrics, in this article our key focus is to discuss the RMT-based approaches to standard high-dimensional problems and their novelties in comparison to traditional methods. The article is organized as follows. In \Cref{sec:RMT}, we discuss the key theoretical results from random matrix theory which provides a framework for the statistical methods. We focus mainly on the asymptotic theory of the spectrum of the two kinds of random matrices - covariance matrix and the ratio of covariance matrices, for their widespread applications in statistics. For each of the two kinds of matrices, we discuss the properties of their bulk spectrum and behavior at the edge of the spectrum, when the matrices are of large dimensions. In the next \Cref{sec:application}, we discuss statistical applications of RMT. We focus on four problems: inference on covariance matrices, application in PCA, application in statistical signal detection removing noise, and changepoint detection.  

\Cref{th:th1} being our contribution, we present the proof of that theorem in the appendix section. We also present a few theoretical applications demonstrating the novelty of this result in \Cref{sebsec: infcovmat}. The rest of the theorems have appeared in the cited papers and we refer to those cited papers for their proof. 

\subsection{Notations and Abbreviations} 
In this paper, $\convP$ means convergence in probability, $\convD$ means convergence in distribution. For a random variable $X$, $F_X(\cdot)$ denotes its CDF. $\mathbf{1}(\cdot)$ denotes the indicator function. RMT means Random Matrix Theory, ESD stands for Empirical SPectral Distribution, LSD indicates Limiting Spectral Distribution. \textit{as} $\mu$ means almost surely wrt the measure $\mu$. 

\section{Background and Motivation}
\label{sec:motivation}

Random matrices play a fundamental role in statistical analysis, particularly in the study of multivariate data. Classical multivariate analysis, as detailed in influential works such as \citet{mardia2024multivariate}, and \citet{muirhead2009aspects}, frequently addresses key problems through the analysis of random matrices. These problems are typically formulated in terms of the eigen-decomposition of Hermitian or symmetric matrices and can be broadly classified into two categories. 

The first category involves the eigen-analysis of a single Hermitian matrix, often referred to as the single Wishart problem, encompassing methods such as principal component analysis (PCA), factor analysis, and tests for population covariance matrices in one-sample problems. The second category includes generalized eigenvalue problems involving two independent Hermitian matrices of the same dimension, commonly known as the double Wishart problem. This includes applications like multivariate analysis of variance (MANOVA), canonical correlation analysis (CCA), tests for equality of covariance matrices, and hypothesis testing in multivariate linear regression.

Beyond these, random matrices also play a natural role in defining and characterizing estimators in multivariate linear regression, classification (involving sample covariance matrices), and clustering (using pairwise distance or similarity matrices). The analysis of eigenvalues and eigenvectors of random symmetric or Hermitian matrices has a long history in statistics, dating back to \citet{pearson1901liii} pioneering work on dimensionality reduction through PCA. This article provides a concise overview of these classical problems to set the stage for the broader discussion of random matrix theory in statistical applications.

Principal component analysis (PCA) is very useful tool for data reduction and model building. The formulation of PCA in classical multivariate analysis at the population level is as follows. Suppose that we measure \( p \) variables (assume real-valued, for simplicity), expressed as a random vector \(\mathbf{X} = \left(X^{(1)}, \ldots, X^{(p)}\right)^{T}\). Suppose also that the random vector \(\mathbf{X}\) has finite variance \(\mathbf{\Sigma} = \mathbb{E}\big[(\mathbf{X} - \mathbb{E}[\mathbf{X}])(\mathbf{X} - \mathbb{E}[\mathbf{X}])^{T}\big]\). The primary goal of PCA is to obtain a lower-dimensional representation of the data in the form of linear transformations of the original variable, subject to the condition that the residual variance is as small as possible. 

This can be achieved by considering a sequence of linear transformations given by \(\mathbf{v}_k^T \mathbf{X}, k = 1, 2, \ldots, p\), satisfying the requirement that \(\mathrm{Var}(\mathbf{v}_k^T \mathbf{X})\) is maximized subject to the conditions that \(\mathbf{v}_k\) are unit norm vectors in \(\mathbb{R}^p\) (or \(\mathbb{C}^p\) if the data is complex-valued), and \(\mathbf{v}_k\) is orthonormal to \(\{\mathbf{v}_j : j = 1, \ldots, k-1\}\), i.e., \(\mathbf{v}_k^T \mathbf{v}_j = 0\) for \(j = 1, \ldots, k-1\). This optimization problem can be solved in terms of the spectral decomposition of the nonnegative definite Hermitian matrix \(\mathbf{\Sigma}\):
\begin{equation}
\mathbf{\Sigma} \mathbf{v}_k = \ell_k \mathbf{v}_k, \quad k = 1, \ldots, p,
\end{equation}
where \(\mathbf{v}_k\) are the orthonormal vectors. Here, \(\ell_k\) (always real-valued) is an eigenvalue associated with \(\mathbf{v}_k\). Note that in this formulation the eigenvalues \(\ell_k\) are ordered, i.e., \(\ell_1 \geq \cdots \geq \ell_p \geq 0\). If \(\ell_k\) is of multiplicity one, then \(\mathbf{v}_k\) is unique up to a sign change. 

In practice, \(\mathbf{\Sigma}\) is unknown and we typically observe a sample \(\mathbf{X}_1, \ldots, \mathbf{X}_n\) for the variable \(\mathbf{X}\). In that case, the empirical version of PCA replaces \(\mathbf{\Sigma}\) by the sample covariance matrix, \(\mathbf{S}_n = (n-1)^{-1} \sum_{i=1}^n (\mathbf{X}_i - \overline{\mathbf{X}})(\mathbf{X}_i - \overline{\mathbf{X}})^T\), and performs the spectral decomposition for \(\mathbf{S}_n\). 

The corresponding eigenvectors \(\widehat{\mathbf{v}}_k\) are often referred to as the sample principal components. The corresponding ordered eigenvalues \(\widehat{\ell}_k\) are typically used to detect the dimension of the reduction subspace. One of the commonly used techniques is to plot the eigenvalues against their indices (so-called "scree plot") and then look for an "elbow" in the plot. 

There are formal tests based on likelihood ratios (see, \citet{mardia2024multivariate}, \citet{muirhead2009aspects}) that assume that, after a certain index, the eigenvalues are all equal and that the observations are Gaussian. Notice that the name "single Wishart" arises from the fact that if \(\mathbf{X}_1, \ldots, \mathbf{X}_n\) are i.i.d. \(N_p(\mathbf{0}, \mathbf{\Sigma})\), then \((n-1)\mathbf{S}_n\) has $W_p(\mathbf{\Sigma},n-1)$ distribution.

Under Gaussianity, one of the commonly used tests for sphericity, i.e., the hypothesis \(H_0 : \mathbf{\Sigma} = \mathbf{I}_p\), is Roy's largest root test (\citet{roy1953heuristic}), which rejects \(H_0\) if \(\widehat{\ell}_1\), the largest eigenvalue of \(\mathbf{S}\), exceeds a threshold determined by the level of significance. If \(H_0 : \mathbf{\Sigma} = \sigma^2 \mathbf{I}_p\) for some unknown \(\sigma^2\), the corresponding generalized likelihood ratio test, under an alternative that assumes \(\mathbf{\Sigma}\) to be a rank-one perturbation of \(\sigma^2 \mathbf{I}_p\), rejects for large values of \(\widehat{\ell}_1 / \big(\sum_{j=2}^p \widehat{\ell}_j\big)\) (\citet{johnson1972analysis}; \citet{nadler2008finite}).

A \textit{factor analysis problem} can be seen as a generalization of PCA in that it assumes a certain signal-plus-noise decomposition of the observation vector \(\mathbf{X}\):
\begin{equation}
\mathbf{X} - \boldsymbol{\mu} = \mathbf{L} \mathbf{f} + \boldsymbol{\varepsilon},
\end{equation}
where \(\mathbf{f}\) is an \(m \times 1\) dimensional random vector, \(\mathbf{L}\) is a \(p \times m\) dimensional nonrandom matrix, \(\mathbf{f}\) and \(\boldsymbol{\varepsilon}\) are uncorrelated, and \(\boldsymbol{\varepsilon}\) has mean \(0\) and variance \(\mathbf{\Psi}\), a \(p \times p\) diagonal matrix. For identifiability, it is typically assumed that \(\mathbb{E}[\mathbf{f}] = 0\) and \(\mathbb{E}[\mathbf{f} \mathbf{f}^T] = \mathbf{I}_m\). 

Under this setting, the covariance matrix of \(\mathbf{X}\) is of the form \(\mathbf{\Sigma} = \mathbf{L} \mathbf{L}^T + \mathbf{\Psi}\). Thus, if \(\mathbf{\Psi}\) is a multiple of the identity, the problem of estimating \(\mathbf{L}\) from data can be formulated in terms of a PCA of the sample covariance matrix. One important distinction between PCA and factor analysis is that, in the latter case, the practitioner implicitly assumes a causal model for the data. In general, factor analysis problems are often solved through a maximum likelihood approach (see \citet{tipping1999probabilistic}). A more enhanced version of the factor analysis model, the so-called dynamic factor model, is used extensively in econometrics, where the factors \(\mathbf{f}\) are taken to be time-dependent (\citet{forni2000tthe}).

A detailed discussion of various versions of the double Wishart eigenproblem, including a summary of the associated distribution theory when the observations are Gaussian, can be found in \citet{johnstone2009sparse}. We first consider the canonical correlation analysis (CCA) problem within this framework. Again, first, we deal with the formulation at the population level. Suppose that real-valued random vectors $\mathbf{X}$ and $\mathbf{Y}$ are jointly observed, where $\mathbf{X}$ is of dimension $p$ and $\mathbf{Y}$ is of dimension $q$. Then a generalization of the notion of correlation between $\mathbf{X}$ and $\mathbf{Y}$ is expressed in terms of the sequence of canonical correlation coefficients defined as
\begin{equation}
\label{eq: rho}
\rho_k = \max_{(\mathbf{u}, \mathbf{v}) \in S_k} |\mathrm{Cor}(\mathbf{u}^\top \mathbf{X}, \mathbf{v}^\top \mathbf{Y})|, \quad k = 1, 2, \dots, \min\{p, q\},
\end{equation}
where
\[
S_k := \{ (\mathbf{u}, \mathbf{v}) \in \mathbb{R}^{p+q} : \mathbf{u}^\top \mathbf{\Sigma}_{\mathbf{XX}} \mathbf{u} = \mathbf{v}^\top \mathbf{\Sigma}_{\mathbf{YY}} \mathbf{v} = 1; \, \mathbf{u}^\top \mathbf{\Sigma}_{\mathbf{XX}} \mathbf{u}_j = \mathbf{v}^\top \mathbf{\Sigma}_{\mathbf{YY}} \mathbf{v}_j = 0, \, j = 1, \dots, k-1 \},
\]
with $\mathbf{\Sigma}_{\mathbf{XX}} = \mathrm{Var}(\mathbf{X})$, $\mathbf{\Sigma}_{\mathbf{YY}} = \mathrm{Var}(\mathbf{Y})$, and $(\mathbf{u}_k, \mathbf{v}_k)$ denoting the pair of vectors for which the maximum in (\ref{eq: rho}) is attained. If $\mathbf{\Sigma}_{\mathbf{XY}} = \mathrm{Cov}(\mathbf{X}, \mathbf{Y})$, then the optimization problem (\ref{eq: rho}) can be formulated as the following generalized eigenvalue problem: the successive canonical correlations $\rho_1 \geq \cdots \geq \rho_{\min\{p,q\}} \geq 0$ satisfy the generalized eigen-equations
\begin{equation}
\label{eq: char poly}
\det (\mathbf{\Sigma}_{\mathbf{XY}} \mathbf{\Sigma}_{\mathbf{YY}}^{-1} \mathbf{\Sigma}_{\mathbf{YX}} - \rho^2 \mathbf{\Sigma}_{\mathbf{XX}}) = 0.
\end{equation}
When we have $n$ samples $\{(\mathbf{X}_i, \mathbf{Y}_i) : i = 1, \dots, n\}$ we can replace $\mathbf{\Sigma}_{\mathbf{XX}}, \mathbf{\Sigma}_{\mathbf{XY}}$ and $\mathbf{\Sigma}_{\mathbf{YY}}$ by their sample counterparts and, assuming $n > \max\{p, q\}$, the corresponding sample canonical correlations $r_1 \geq \cdots \geq r_{\min\{p,q\}} \geq 0$ satisfy the sample version of \Cref{eq: char poly}. It is shown in \citet{mardia2024multivariate} that in the latter case, we can reformulate the corresponding generalized eigenanalysis problem as solving
\begin{equation}
\det (\mathbf{U} - r^2 (\mathbf{U} + \mathbf{V})) = 0,
\end{equation}
where $\mathbf{U}$ and $\mathbf{V}$ are independent Wishart matrices if $(\mathbf{X}_i, \mathbf{Y}_i)$ are i.i.d. Gaussian and $\mathbf{\Sigma}_{\mathbf{XY}} = \mathbf{0}$, i.e., $\mathbf{X}$ and $\mathbf{Y}$ are independently distributed.

Next, we consider the multivariate Linear Regression model, 
\begin{equation}
\label{eq: High dim Lin Reg}
    \mathbf{Y} = \mathbf{XB} + \mathbf{E}
\end{equation}
where $\mathbf{Y} = [Y_1: \cdots: Y_m] \in \R^{n \times m}$ is the response matrix consisting $n$ observations for each of the $m$ response variable, $\mathbf{X} \in \R^{n \times p}$ is the design matrix where $p$ is the number of covariates. $\mathbf{E}$ denotes the error matrix. For inference purposes, it is further assumed that $\mathbf{E} \sim NDM(0,\mathbf{\Sigma})$ i.e. rows of $\mathbf{E}$ are iid from $N(0,\mathbf{\Sigma})$.  Then, as described in \citet{mardia2024multivariate}, the union-intersection test for the linear hypothesis of the form $H_0 : \mathbf{C}\mathbf{B}\mathbf{D} = \mathbf{0}$ where $\mathbf{C}$ and $\mathbf{D}$ are specified conformable matrices, can be expressed in terms of the largest eigenvalue of $\mathbf{U} (\mathbf{U} + \mathbf{V})^{-1}$ where $\mathbf{U}$ and $\mathbf{V}$ are appropriately specified independent Wishart matrices (under Gaussianity of the entries of $\mathbf{E}$).

The two-sample test for equality of variances assumes that we have i.i.d. samples from two normal populations $N_p(\boldsymbol{\mu}_1, \mathbf{\Sigma}_1)$ and $N_p(\boldsymbol{\mu}_2, \mathbf{\Sigma}_2)$ of sizes $n_1$ and $n_2$, say. Then several tests for the hypothesis $H_0 : \mathbf{\Sigma}_1 = \mathbf{\Sigma}_2$ can be formulated in terms of functionals of the eigenvalues of $\mathbf{U} (\mathbf{U} + \mathbf{V})^{-1}$ where $\mathbf{U} = (n_1 - 1)\mathbf{S}_1$ and $\mathbf{V} = (n_2 - 1)\mathbf{S}_2$ are the sample covariances for the two samples, which would follow independent Wishart distributions in $p$ dimensions with d.f. $n_1 - 1$ and $n_2 - 1$ and dispersion matrix $\mathbf{\Sigma}_1 = \mathbf{\Sigma}_2$ under $H_0$.

Now it is to be noted that the traditional methods to deal with the above problems assume the dimension of the data to be fixed and relatively small compared to the number of data points. But in the modern era, most of the high-dimensional data arising in fields such as genomics, economics, atmospheric
 science, chemometrics, and astronomy, to name a few, are of enormously large dimensions which makes it very challenging to apply the traditional methods directly to those datasets. And so, to accommodate the analysis of such datasets, it is imperative to either modify or reformulate some of the statistical techniques. This is where RMT has been playing a significant role, especially over the last decade. In the next sections, we develop these modern RMT-based methods with a key focus on their applications in statistics.

\section{High Dimensional Random Matrices}
\label{sec:RMT}

In Random Matrix Theory, two particular kinds of random matrices draw special attention for their remarkable application in statistics - Covariance Matrices and F-type Matrices. In this section, we discuss the theoretical properties of these two kinds of random matrices, which play a key role in most modern developments in high-dimensional statistics. Most of these results focus on the spectrum's behavior when the matrix has a large dimension. Hence to address those properties, we first give a basic introduction to these matrix models and describe a couple of key questions associated with it. 

In classical random matrix theory, in the context of covariance matrices, one of the most fundamental and crucially studied matrices is the Wishart Matrix. The Wishart matrix is defined as specifying two sequences of integers $n$, the sample size, and $p = p(n)$, the data dimension. Most of the results additionally assume that the sequences are related so that, as $n \to \infty$, $p = p(n) \to \infty$ satisfying, 
\begin{equation*}
    \underset{n \to \infty}{\text{lim}} \frac{p}{n} = \gamma \in (0,\infty) 
\end{equation*}
So if, $X_1,\cdots,X_n \overset{iid}{\sim} N_p(\mathbf{0},\mathbf{\Sigma})$, and $\mathbf{X} = [X_1:\cdots:X_n]$, then the distribution of $\mathbf{X}\mathbf{X^T}$ is called Wishart distribution with parameter $\mathbf{\Sigma}$, degree of freedom $n$ and dimension $p$ and abbreviated as $W_p(\mathbf{\Sigma},m)$. A density of the distribution is given by 
\[
f_W(\mathbf{X}) = \frac{|\mathbf{X}|^{(n-p-1)/2} e^{-\frac{1}{2} \text{tr}(\Sigma^{-1} \mathbf{X})}}{2^{np/2} |\Sigma|^{n/2} \Gamma_p\left(\frac{n}{2}\right)}
\]
where $\Gamma_p(\cdot)$ is the multivariate gamma function. The distribution was first studied by \citet{wishart1928generalised} and continues to be a fundamental focus in multivariate statistics thereafter. The central reason for Wishart matrices being so frequent and useful in statistics is their association with sample covariance matrices - the sample covariance matrix of a normal random sample follows a Wishart distribution i.e. If the data $X_1,\cdots,X_n \overset{iid}{\sim} N_p(\mathbf{0},\mathbf{\Sigma})$, and $\mathbf{S} = \frac{1}{n}\sum_{i=1}^n (X_i - \overline{X})(X_i - \overline{X})^T$ is the sample covariance matrix, then $n\mathbf{S} \sim W_p(\mathbf{\Sigma},n-1)$. Further detailed properties of the distribution can be found in \citet{muirhead2009aspects},  \citet{mardia2024multivariate}. From the definition, it can be seen that the Wishart distribution is the analog of the chi-square distribution of the univariate case. Similarly, the univariate $F-$distribution has also an extension for matrices, which is called Matrix $F-$distribution. If $\mathbf{A} \sim W_p(I_p,\nu)$ and $\mathbf{B}\sim W_p(I_p,\delta)$ and $\mathbf{A,B}$ are independent then the distribution of $\mathbf{B}^{-1/2}\mathbf{A}\mathbf{B}^{-1/2}$ is called a matrix $\mathbf{F}(I_p,\nu,\delta)$ distribution. The distribution was originally derived by \citet{olkin1964multivariate}. \citet{perlman1977note} discusses several interesting properties of this distribution. Usually if $\mathbf{A,B}$ are independent random matrices such that $\mathbf{A} \sim W_p(\mathbf{\Sigma},m)$ and $\mathbf{B} \sim W_p(\mathbf{\Sigma},n)$ and $\Sigma$ is positive definite and $m \geqslant p$, then $\mathbf{A}^{-1}\mathbf{B}$ is called an $F-$type matrix in the literature. For their widespread applications such as in Linear Discriminant Analysis, Canonical Correlation Analysis, etc, $F-$type matrices are also extensively studied. However, the properties of these matrices, which played a crucial role in statistical inference and related fields over decades, have been studied beyond the parametric framework, under minimalistic assumptions mostly for the high-dimensional setup. We discuss the properties of those matrices in terms of their spectrum in both parametric and nonparametric frameworks.

\subsection{Spectral Properties of large Sample Covariance Matrices}
\label{subsec:cov matrices}
The sample covariance matrix is one of the most important random matrices in multivariate statistical inference. It is fundamental in hypothesis testing, principal component analysis, factor analysis, and discrimination analysis. Many test statistics are defined by their eigenvalues. However, for large matrices, it is more convenient to study the asymptotic behavior of their spectrum as they exhibit nice properties. 
\subsubsection{Properties of the whole Spectrum}
\label{subsubsec: spectrum}

Suppose $\mathbf{X}$ is a $n \times n$ random matrix having eignevalues $\lambda_1,\cdots,\lambda_n \in \mathbb{C}$. Then the empirical distribution of the eigenvalues of $\mathbf{X}$ is called the empirical spectral distribution(ESD) of $\mathbf{X}$. If the matrix $\mathbf{X}$ is real, symmetric then all of its eigenvalues are real and hence the empirical spectral distribution is given by $\hat{F}(x) = \frac{1}{n} \sum_{i=1}^n \mathbf{1}(\lambda_i \leqslant x)$ which is the case for Wishart matrices. In random matrix theory, the ESD is crucial to study as most of the properties of a matrix can be reformulated in terms of its eigenvalues and hence is of key interest. In different domains like machine learning, signal processing, and wireless communications, several functions of the eigenvalues give important objects to study (eg. \citet{tulino2004random}, \citet{couillet2022random} etc). 

In the parametric setup, under the normality assumption of the data, when we have the exact distribution of the sample covariance matrix, one of the most fundamental questions one can ask is how to characterize the joint distribution of the spectrum. If $\mathbf{X} \sim W_p(\mathbf{\Sigma},n)$, then the rank of $\mathbf{X}$ is min$\{p,n\}$ almost surely. Now for $n \leqslant p-1$, the rank of $\mathbf{X} \leqslant p - 1$ and hence the eigenvalues do not have a joint pdf. However, for $n > p-1$, the joint probability density function (pdf) for the eigenvalues of $\mathbf{X}$ exists and can be found in \citet{muirhead2009aspects} and a detailed study on their joint distribution can be found in \citet{james2014concise}. Furthermore, for $n > p-1$, the following central limit theorem holds for log-transformed eigenvalues of $\mathbf{X}$. 

\begin{theorem}
\label{th:th1}
Let $\mathbf{X}_n \sim W_p(\mathbf{\Sigma},n)$ where $n \geqslant p$ and $\mathbf{\Sigma}$ is positive definite. Let $\lambda_1^{(n)},\cdots,\lambda_p^{(n)}$ be the eigenvalues of $\mathbf{X}_n$ and $\lambda_1,\cdots,\lambda_p$ be the eigenvalues of $\mathbf{\Sigma}$. Then, 
\begin{equation*}
    \underset{x \in \mathbb{R}}{\text{sup}} \hspace{0.1cm} \left| P \left( \sqrt{\frac{n}{2p}} \left( \sum_{i=1}^p \text{log}\left( \frac{\lambda_i^{(n)}}{\lambda_i}\right) - \sum_{i=1}^p\text{log} \hspace{0.1cm} (n-p+i) \right) \leqslant x \right) - \Phi(x) \right| = O\left(\frac{p}{\sqrt{n}} \right)
\end{equation*}
where $\Phi(x)$ denotes  the Standard Normal CDF.
\end{theorem}

The proof of \Cref{th:th1} can be found in the appendix. Hence, for $\frac{p}{\sqrt{n}} \to 0$ as $n \to \infty$, we have 

\begin{equation*}
    \sqrt{\frac{n}{2p}} \left( \sum_{i=1}^p \text{log}\left( \frac{\lambda_i^{(n)}}{\lambda_i}\right) - \sum_{i=1}^p\text{log} \hspace{0.1cm} (n-p+i) \right) \convD N(0,1)
\end{equation*}
It is to be noted that the above central limit theorem is very useful for one-sample and two-sample testing for covariance matrices, to approximate the power function of such tests, and also for inference on covariance matrices in high dimensional linear models. A detailed discussion of these applications can be found in \Cref{sebsec: infcovmat}. The theorem also provides a rate of convergence of the eigenvalues of the sample covariance matrix to the population covariance matrix. Often for approximation purposes, functions of the spectrum of the population covariance matrices are estimated using that of the corresponding sample covariance matrix. In such cases, the theorem provides an upper bound of the error. Hence, it is useful for sample size determination if there is a predetermined allowable upper bound on the error. 

In this context, a natural follow-up question is whether the weak limit of the ESD for Wishart matrices exists. The celebrated \textit{Marcenko-Pastur Law} answers this question in the context of sample covariance matrices. With an assumption of the finiteness of the fourth moment of the entries of the data matrix, \citet{marchenko1967distribution} showed that depending on the value of $\gamma = \underset{n \to \infty}{\text{lim}} \frac{p}{n}$, the weak limit of the ESD of sample covariance matrices exist.

\begin{theorem}[Marcenko-Pastur Law]
\label{th: th2}
    Suppose that \( \mathbf{X} \) is a \( p \times n \) matrix with i.i.d. real- or complex-valued entries with mean 0 and variance 1. Suppose $\underset{n \to \infty}{\text{lim}} \frac{p}{n} = \gamma \in (0,\infty) $ . Then, as \( n \to \infty \), the empirical spectral distribution (ESD) of \( \mathbf{S} = \frac{1}{n}\mathbf{XX}^T \) converges almost surely in distribution to a nonrandom distribution, known as the Marcenko–Pastur law and denoted by \( F_\gamma \). If \( \gamma \in (0, 1] \), then \( F_\gamma \) has the p.d.f.:
\begin{equation}
    f_\gamma(x) = \frac{\sqrt{(b_+(\gamma) - x)(x - b_-(\gamma))}}{2\pi \gamma x}, \quad b_-(\gamma) \leq x \leq b_+(\gamma),
\end{equation}
where
\[
b_\pm(\gamma) = \left(1 \pm \sqrt{\gamma}\right)^2.
\]
For \( x \) outside this interval, \( f_\gamma(x) = 0 \).

If \( \gamma \in (1, \infty) \), then \( F_\gamma \) is a mixture of a point mass at 0 and the p.d.f. \( f_{1/\gamma}(x) \), with weights \( 1 - 1/\gamma \) and \( 1/\gamma \), respectively.

\end{theorem}

\begin{figure}
    \centering
    \includegraphics[width=0.5\linewidth]{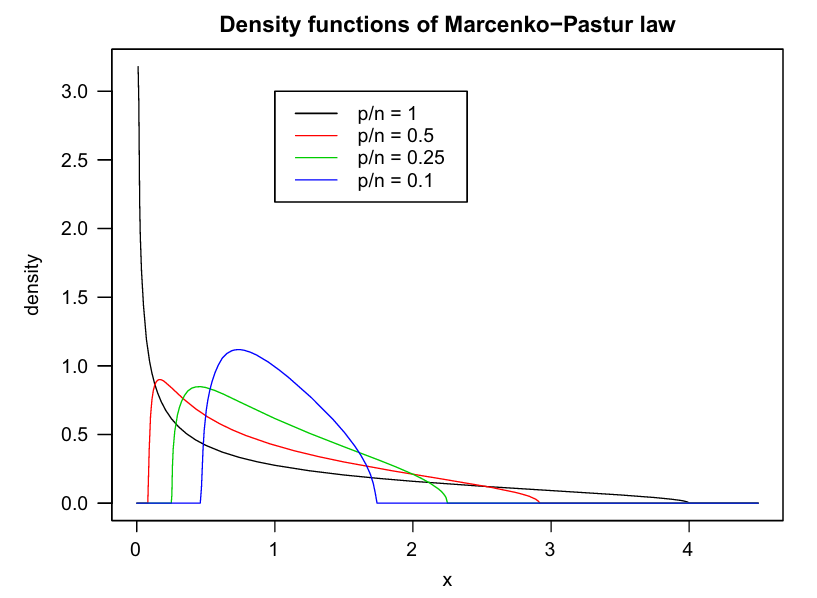}
    \caption{Marcenko–Pastur density functions for $\gamma = 0.1,0.25,0.5,1$}
    \label{fig:marcenko pastur law}
\end{figure}

 It is to be noted that the above result is distribution-free, in the sense the limiting distribution only depends on the limiting ratio of data dimension and sample size ($\gamma$) and is free of the data distribution. As $\gamma$ increases from $0$ to $1$, the spread of the eigenvalues also increases. However, a necessary condition for the weak limit of the ESD to exist is to $\gamma > 0$. For $\gamma = 0$, as illustrated in \Cref{fig:marcenko pastur law}, the maximum and minimum eigenvalues converge to $1$ and hence Marcenko-Pastur law does not hold for this case. However, with an assumption of the finiteness of the fourth moment of the entries of $\mathbf{X}$, applying a suitable centering and scaling to the matrix $\mathbf{S}$, \citet{bai1988convergence} derived the weak limit of the ESD of the transformed matrix when $\frac{p}{n} \to 0$. 

 \begin{theorem}[\citet{bai1988convergence}] 
 \label{th: th3}
 Suppose that \( \mathbf{X} \) is a \( p \times n \) matrix with i.i.d. real-valued entries with mean 0 and variance 1 with finite fourth moment. Suppose $\underset{n \to \infty}{\text{lim}} \frac{p}{n} = 0 $ . Then, as \( p \to \infty \), the empirical spectral distribution (ESD) of \( \mathbf{S}_p = \frac{1}{2\sqrt{np}} \left(\mathbf{XX}^T - n I_p \right) \) converges almost surely in distribution to a nonrandom distribution, known as semi-circular distribution having the pdf
 \begin{equation}
f(x) = \frac{1}{2\pi} \sqrt{4 - x^2}, \quad -2 \leq x \leq 2
\end{equation}

 \end{theorem}

 Like the Marcenko-Pastur law, the above theorem is also distribution-free and provides the rate at which the eigenvalues of $\mathbf{S} = \frac{1}{n}\mathbf{XX}^T$ goes to 1 when $\frac{p}{n} \to 0$. However, one potential disadvantage of the above two theorems is the requirement of the i.i.d data. In both of the theorem, in the data matrix $\mathbf{X} = [X_1,\cdots,X_n]$, where each of the columns ($X_i$) represents a data point of dimension $p$, though each of the columns can be assumed to be independent, in practice it is very unlikely that the entries of a single data point in $\mathbb{R}^p$ will be mutually independent as well. Henceforth substantial progress has been made to generalize these results by relaxing the conditions of independence within the columns. If $Y_1,\cdots,Y_n$ are i.i.d. $p-$dimensional data points with covariance matrix $\mathbf{\Sigma}$, then $\mathbf{Y} = [Y_1,\cdots,Y_n] = \mathbf{\Sigma}^{\frac{1}{2}} \mathbf{X}$, where $\mathbf{X}$ satisfies the conditions of \Cref{th: th2} and \ref{th: th3}. In this context, \citet{silverstein1995strong} develops the Marcenko-Pastur law for the ESD of $\frac{1}{n} \mathbf{\Sigma}^{\frac{1}{2}}\mathbf{X}\mathbf{X}^T\mathbf{\Sigma}^{\frac{1}{2}}$ when $\frac{p}{n} \to \gamma \in (0,\infty)$ under the same conditions as of \Cref{th: th2}. For $\frac{p}{n} \to 0$, under the conditions of \Cref{th: th3}, \citet{bao2012strong} showed the ESD of $\sqrt{\frac{n}{p}} \left( \frac{1}{n} \mathbf{\Sigma}^{\frac{1}{2}}\mathbf{X}\mathbf{X}^T\mathbf{\Sigma}^{\frac{1}{2}} - \mathbf{\Sigma}  \right) = \sqrt{\frac{n}{p}} \mathbf{\Sigma}^{\frac{1}{2}} \left( \frac{1}{n} \mathbf{X}\mathbf{X}^T - \mathbf{I} \right) \mathbf{\Sigma}^{\frac{1}{2}}$ converges almost surely in distribution to a nonrandom distribution.

 Further research has been done to develop similar results under different forms of dependence. For instance, \citet{yin1987limit} derived similar results when $X_1,\cdots,X_n$ are i.i.d from a spherically symmetric distribution. \citet{hui2010limiting}, \citet{wei2016limiting} considered the case when the data points come from a $m-$dependent process. \citet{hofmann2008wigner} and \citet{friesen2013gaussian} assumed that the entries of the data matrix $\mathbf{X} = [X_1,\cdots,X_n]$ can be
 partitioned into independent subsets while allowing the entries from the same subset to be
 dependent. \citet{gotze2006limit} replaced the independent assumption by a technical martingale-type condition. \citet{yao2012note} develops a version of the Marcenko-Pastur law when $X_1,\cdots,X_n$ are independent and entries of each $X_i$ comes from a linear time series process.

 \subsubsection{Properties of extreme eigenvalues}
 \label{subsubsec: extreme eval}

In the previous section, we have a detailed description of the limit of the ESD of random matrices. However, in many situations, it is important to know whether the sample eigenvalues of $\mathbf{S}$ (as in \Cref{th: th2}) lie inside the support of $F_\gamma$ as well. For instance, in signal processing, pattern recognition, edge detection, and many other areas, the support of the LSD of the population covariance matrices consists of several disjoint pieces. So it is essential to know whether or not the LSD of the sample covariance matrices is also separated
 into the same number of disjoint pieces, and under what conditions this is true. Also, many statistics can be written as a function of the integrals of the ESD of the random matrix. For example, the determinant of the sample covariance matrix is very useful in wireless communication and signal processing (\citet{paul2014random}) which can be written as 
 \begin{equation}
 \label{eq:det}
     \det(\mathbf{A}) = \prod_{j=1}^{n} \lambda_j = \exp \left( n \int_0^{\infty} \log x \, F^{\mathbf{A}}(dx) \right)
 \end{equation}
So under the knowledge of the asymptotic distribution of the ESD, usually the Helly-Bray theorem (\citet{billingsley2013convergence}) is used to obtain an approximation of the statistic. But often such functions are not bounded (e.g. the function in \ref{eq:det} is $log x$ which is unbounded. As a result, the LSD and Helly-Bray theorem cannot be used to approximate the statistics. This limitation reduces the usefulness of the LSD. However, in many cases, the supports of the LSDs are compact intervals. Still, this alone does not guarantee that the Helly-Bray theorem can be applied unless one also proves in addition that the extreme eigenvalues of the random matrix stay within certain bounded intervals. These examples demonstrate that knowledge about the weak limit of the ESD is not sufficient. Furthermore, extreme eigenvalues of random matrices themselves occur naturally in many problems such as principal component analysis. Henceforth studies regarding the asymptotic properties of the extreme eigenvalues of random matrices are extremely important. Under the assumption of the finiteness of the fourth moment of the i.i.d. entries, \citet{yin1984limit} proved that the maximum eigenvalue of $\mathbf{S}$ (as in \Cref{th: th2}) converges almost surely. 

\begin{theorem}[\citet{yin1984limit}] 
\label{th:th 4}
Suppose that \( \mathbf{X} \) is a \( p \times n \) matrix with i.i.d. real-valued entries with mean 0 and variance $\sigma^2$ and finite fourth moment. Suppose $\underset{n \to \infty}{\text{lim}} \frac{p}{n} = \gamma \in (0,\infty) $ . Suppose $\lambda_{\text{max}}(n)$ is the maximum eigenvalue of the $p \times p$ random matrix \( \mathbf{S} = \frac{1}{n}\mathbf{XX}^T \). Then  
\begin{equation}
    \underset{n\to \infty}{\text{lim}} \lambda_{\text{max}}(n) = (1 + \sqrt{\gamma})^2 \sigma^2 \hspace{0.3cm} \text{a.s.} 
\end{equation}
    
\end{theorem}

A similar result was developed in \citet{bai2008limit} for the smallest eigenvalue of $\mathbf{S}$ as well under the same set of assumptions as of \Cref{th:th 4} as well when $p < n$.

\begin{theorem}[\citet{bai2008limit}] 
\label{th:th 4}
Suppose that \( \mathbf{X} \) is a \( p \times n \) matrix with i.i.d. real-valued entries with mean 0 and variance $\sigma^2$ and finite fourth moment. Suppose $\underset{n \to \infty}{\text{lim}} \frac{p}{n} = \gamma \in (0,1) $ . Suppose $\lambda_{\text{min}}(n)$ is the smallest eigenvalue of the $p \times p$ random matrix \( \mathbf{S} = \frac{1}{n}\mathbf{XX}^T \). Then  
\begin{equation}
    \underset{n\to \infty}{\text{lim}} \lambda_{\text{min}}(n) = (1 - \sqrt{\gamma})^2 \sigma^2 \hspace{0.3cm} \text{a.s.} 
\end{equation}
    
\end{theorem}

These results give an accurate idea of the asymptotic range of the eigenvalues of sample covariance matrices under very mild assumptions. However many classical tests in multivariate analysis consist of the largest eigenvalues of sample covariance matrices (eg. Roy's largest root test) which makes the asymptotic distributions of the maximum eigenvalue of special interest. In the celebrated paper \citet{johnstone2001distribution}, the limiting distribution of the largest eigenvalue of the sample covariance matrix was derived when the entries of the data matrix are i.i.d from the standard normal distribution. Suppose that \( \mathbf{X} = ((X_{ij})) \) is an \( p \times n \) matrix with entries are i.i.d. from standard normal distribution, 
\[
X_{ij} \sim N(0, 1).
\]
Let $l_1$ be the largest sample eigenvalue of the Wishart matrix \( \mathbf{X}\mathbf{X}^T \). Define the centering and scaling constants as follows:
\begin{equation}
\label{eq:mu}
    \mu_{n,p} = \left( \sqrt{n-1} + \sqrt{p} \right)^2
\end{equation}
\begin{equation}
\label{eq:sigma}
    \sigma_{n,p} = \left( \sqrt{n-1} + \sqrt{p} \right) \left( \frac{1}{\sqrt{n-1}} + \frac{1}{\sqrt{p}} \right)^{1/3}
\end{equation}
The Tracy-Widom law of order 1 has the distribution function defined by:
\begin{equation}
\label{eq:TW cdf}
    F_1(s) = \exp \left( -\frac{1}{2} \int_s^\infty \left[ q(x) + (x - s) q^2(x) \right] dx \right) , \hspace{0.2cm} s \in \mathbb{R}
\end{equation}
where \( q(x) \) solves the nonlinear Painlevé II differential equation:
\begin{equation}
    q''(x) = x q(x) + 2 q^3(x)
\end{equation}
with the asymptotic condition:
\begin{equation}
    q(x) \sim \text{Ai}(x) \quad \text{as} \quad x \to +\infty,
\end{equation}
where \( \text{Ai}(x) \) denotes the Airy function. This distribution was found by \citet{tracy1996orthogonal} as the limiting law of the largest eigenvalue of an \( n \times n \) Gaussian symmetric matrix. In terms of these distributions, the asymptotic distribution of $l_1$ can be stated as follows,

\begin{theorem}[\citet{johnstone2001distribution}]
\label{th: tracy widom}
    Suppose that \( \mathbf{X} = ((X_{ij})) \) is an \( p \times n \) matrix whose entries are i.i.d. from standard normal distribution i.e. $X_{ij} \overset{i.i.d.}{\sim} N(0, 1)$. If $\frac{p}{n} \to \gamma \in (0,\infty)$, and $l_1$ denotes the highest eigenvalue of $\mathbf{X}\mathbf{X}^T$ then, 
    \begin{equation}
        \frac{l_1 - \mu_{n,p}}{\sigma_{n,p}} \convD W_1 \sim F_1
    \end{equation}
where $\mu_{n,p},\sigma_{n,p}$ are as in \ref{eq:mu} and \ref{eq:sigma} respectively and $W_1$ is a random variable following Tracy-Widom distribution defined in \ref{eq:TW cdf}. 
\end{theorem}

In \citet{karoui2003largest}, the same result was extended for the cases $\gamma=0,\infty$ as well. 
\begin{figure}
    \centering
    \includegraphics[width=0.6\linewidth]{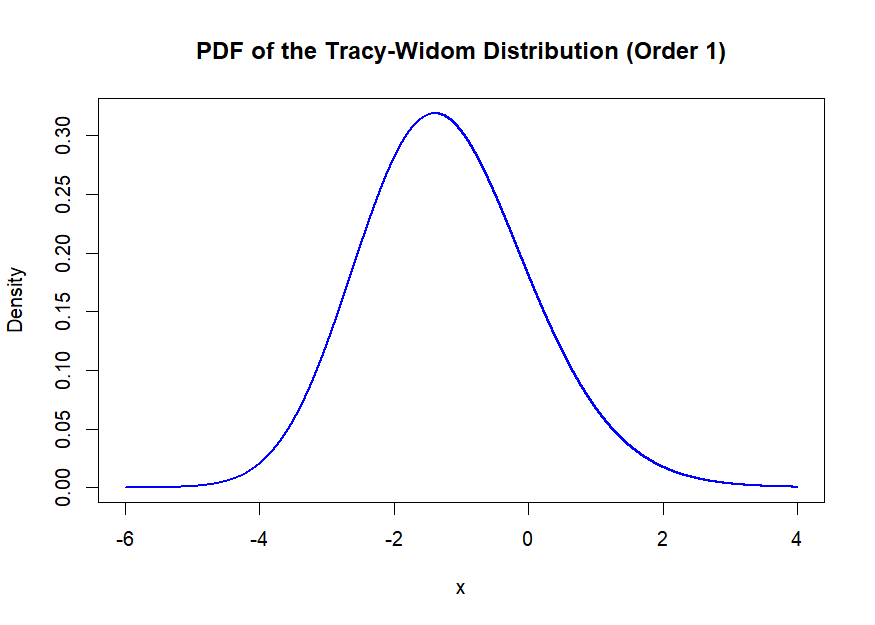}
    \caption{Density function for the Tracy-Widom Distribution}
    \label{fig:enter-label}
\end{figure}
These results are very useful in single Wishart (e.g. principal
 component analysis (PCA), factor analysis, and tests for population covariance matrices in one-sample problems) and double Wishart problems (e.g. multivariate analysis of variance (MANOVA), canonical correlation analysis (CCA), tests for equality of
 covariance matrices and tests for linear hypotheses in multivariate linear regression problems). Many asymptotic generalizations of classical tests (e.g. Roy's largest root test) have been obtained using the above results which have a wide range of applications in signal processing, wireless communication (\citet{paul2014random}), and machine learning (\citet{couillet2022random}). \Cref{sec:application} has a detailed discussion on these applications.

To check the practical applicability of \Cref{th: tracy widom}, for the purpose of approximation, a simulation study was done (\citet{johnstone2001distribution}). First, for square cases $n = p = 5, 10$ and 100, using $R = 10,000$ replications, results are shown in \Cref{table: table tracy widom}. Even for $5 \times 5$ and $10 \times 10$, the approximation seems to be quite good in the right-hand tail at conventional significance levels of 10\%, 5\%, and 1\%. At $100 \times 100$, the approximation seems reasonable throughout the range. The same general picture holds for $n/p$ in the ratio 4:1. Even for $5 \times 20$ matrices, the approximation is reasonable, if not excellent, at the conventional upper significance levels.

A further summary message from these computations is that in the null Wishart case, about 80\% of the distribution lies below $\mu_{np}$ and 95\% below $\mu_{np}+ \sigma_{np}$. \citet{ma2012accuracy} has a detailed discussion on the accuracy of the Tracy-Widom laws. 

\begin{table}[ht]
    \centering
    \caption{(\citet{johnstone2001distribution}) Simulations for finite \( n \times p \) versus Tracy–Widom Limit. The first column shows the probabilities of the \( F_1 \) limit distribution corresponding to fractions in the second column. The next three columns show estimated cumulative probabilities for \( l_1 \), centered and scaled as in \Cref{eq:mu} and \ref{eq:sigma}, in \( R = 10,000 \) repeated draws from \( W_p(n, I) \) with \( n = p = 5, 10, 100 \). The following three cases have \( n:p \) in the ratio 4:1. The final column gives approximate standard errors based on binomial sampling. The bold font highlights some conventional significance levels. The Tracy–Widom distribution \( F_1 \) was evaluated on a grid of 121 points \( -6(0.1)6 \) using the Mathematica package \texttt{p2Num} written by Craig Tracy. The remaining computations were done in MATLAB, with percentiles obtained by inverse interpolation and using \texttt{randn()} for normal variates and \texttt{norm()} to evaluate the largest singular values.}
    \label{table: table tracy widom}
    \hspace{0.2cm}
    \begin{tabular}{cccccccccc}
\toprule
Percentile & TW & 5 $\times$ 5 & 10 $\times$ 10 & 100 $\times$ 100 & 5 $\times$ 20 & 10 $\times$ 40 & 100 $\times$ 400 & 2 $\times$ SE \\
\midrule
-3.90 & 0.01 & 0.000 & 0.001 & 0.007 & 0.002 & 0.003 & 0.010 & (0.002) \\
-3.18 & 0.05 & 0.003 & 0.015 & 0.042 & 0.029 & 0.039 & 0.049 & (0.004) \\
-2.78 & 0.10 & 0.019 & 0.049 & 0.089 & 0.075 & 0.089 & 0.102 & (0.006) \\
-1.91 & 0.30 & 0.211 & 0.251 & 0.299 & 0.304 & 0.307 & 0.303 & (0.009) \\
-1.27 & 0.50 & 0.458 & 0.480 & 0.500 & 0.539 & 0.524 & 0.508 & (0.010) \\
-0.59 & 0.70 & 0.697 & 0.707 & 0.703 & 0.739 & 0.733 & 0.714 & (0.009) \\
0.45 & 0.90 & 0.901 & 0.907 & 0.903 & 0.919 & 0.918 & 0.908 & (0.006) \\
0.98 & 0.95 & 0.948 & 0.954 & 0.950 & 0.960 & 0.961 & 0.957 & (0.004) \\
2.02 & 0.99 & 0.988 & 0.991 & 0.991 & 0.992 & 0.993 & 0.992 & (0.002) \\
\bottomrule
\end{tabular}
\end{table}

 \subsection{Asymptotic Properties of $F-$type matrices}
 \label{subsec: Fmatrices}

In this section, we discuss the asymptotic properties of a multivariate $F-$ matrix. Multivariate F-distribution plays a crucial role in several areas of multivariate data analysis, especially when the relationships between multiple variables are tested simultaneously. It has primary application in two-sample tests on covariance matrices, MANOVA (multivariate analysis of variance), multivariate linear regression, and in Canonical Correlation Analysis. Pioneering work by \citet{wachter1980limiting} examined the limiting distribution of the solutions to the equation:
\[
\det\left( \mathbf{X}_{1,n_1} \mathbf{X}_{1,n_1}^T - \lambda \mathbf{X}_{2,n_2} \mathbf{X}_{2,n_2}^T \right) = 0,
\]
where \( \mathbf{X}_{j,n_j} \) is a \( p \times n_j \) matrix with i.i.d. entries from \( N(0,1) \), and \( \mathbf{X}_{1,n_1} \) is independent of \( \mathbf{X}_{2,n_2} \). When \( \mathbf{X}_{2,n_2} \mathbf{X}_{2,n_2}^T \) is of full rank, the solutions to this equation are \( \frac{n_2}{n_1} \) times the eigenvalues of the multivariate F-matrix:
\[
\left( \frac{1}{n_1} \mathbf{X}_{1,n_1} \mathbf{X}_{1,n_1}^T \right) \left( \frac{1}{n_2} \mathbf{X}_{2,n_2} \mathbf{X}_{2,n_2}^T \right)^{-1}.
\]

\citet{yin1983limit} proved the existence of the limiting spectral distribution (LSD) of the matrix sequence \( \{ \mathbf{S}_n \mathbf{T}_n \} \), where \( \mathbf{S}_n \) is a standard Wishart matrix of dimension \( p \) with \( n \) degrees of freedom, and \( \frac{p}{n} \to \gamma \in (0, \infty) \), \( \mathbf{T}_n \) is a positive definite matrix with \( \beta_k(\mathbf{T}_n) \to \mathbf{H}_k \), and the sequence \( \mathbf{H}_k \) satisfies the Carleman condition. In \citet{yin1986limiting}, this result was extended to the case where the sample covariance matrix is formed from i.i.d. real random variables with mean zero and variance one. Building on the work of \citet{yin1983limit}, later \citet{yin1983limiting} demonstrated the existence of the LSD of the multivariate F-matrix. The explicit form of the LSD for multivariate F-matrices was derived by \citet{bai1988limiting} and \citet{silverstein1995strong} and is given by the following theorem.

\begin{theorem}[\citet{bai1988limiting}]
\label{th:th7}
    Let \( \mathbf{F} = \mathbf{S}_{n_1} \mathbf{S}_{n_2}^{-1} \), where \( \mathbf{S}_{n_i} \) (for \( i = 1,2 \)) is a sample covariance matrix with dimension \( p \) and sample size \( n_i \), and the underlying distribution has mean 0 and variance 1. If \( \mathbf{S}_{n_1} \) and \( \mathbf{S}_{n_2} \) are independent, \( \frac{p}{n_1} \to \gamma \in (0,\infty) \), and \( \frac{p}{n_2} \to \gamma' \in (0,1) \), then the limiting spectral distribution (LSD) \( F_{\gamma,\gamma'} \) of \( \mathbf{F} \) exists and has a density function given by:

\[
f_{\gamma,\gamma'}(x) = 
\begin{cases} 
\frac{(1 - \gamma')\sqrt{(b - x)(x - a)}}{2 \pi x (\gamma + x\gamma')}, & \text{if } a < x < b, \\
0, & \text{otherwise},
\end{cases}
\]
where
\[
a = \left( \frac{1 - \sqrt{\gamma + \gamma' - \gamma\gamma'}}{1 - \gamma'} \right)^2, b = \left( \frac{1 + \sqrt{\gamma + \gamma' - \gamma\gamma'}}{1 - \gamma'} \right)^2.
\]

Further, if \( \gamma > 1 \), then \( F_{\gamma,\gamma'} \) has a point mass \( 1 - \frac{1}{\gamma} \) at the origin.
\end{theorem}

Besides the entire ESD, the extreme eigenvalues of multivariate $F-$matrices are also immensely important in many high-dimensional problems such as testing sphericality in covariance matrices, testing equality of multiple covariance matrices, correlated noise detection, etc (\citet{han2016tracy}). The following result from the phenomenal work of \citet{han2016tracy} obtains the limiting distribution of the generalized $F-$type matrices under mild assumptions. Before starting the actual theorem, we first state a condition the data matrix needs to satisfy. 

\begin{definition}
\label{def:def 1}
A real random matrix $\mathbf{Z}$ is said to satisfy \textbf{Condition 1}, if it consists of entries \( \{ Z_{ij} \} \) where \( \{ Z_{ij} \} \) are independent random variables with \( \mathbb{E}[Z_{ij}] = 0 \) and \( \mathbb{E}[|Z_{ij}|^2] = 1 \) and for all \( k \in \mathbb{N} \), there exists a constant \( C_k \) such that \( \mathbb{E}[|Z_{ij}|^k] \leq C_k \).
\end{definition}

In conjunction with the above definition, the following theorem presents the desired limiting distribution. 
\begin{theorem}[\citet{han2016tracy}]
\label{th: tw for Fmat}
     Also assume that the real random matrices \( \mathbf{X} = (X_{ij})_{p \times n} \) and \( \mathbf{Y} = (Y_{ij})_{p \times m} \) are independent and satisfies \textbf{Condition 1}. Set \( m = m(p) \) and \( n = n(p) \). Suppose that
\[
\lim_{p \to \infty} \frac{p}{m} = d_1 > 0, \quad \lim_{p \to \infty} \frac{p}{n} = d_2 > 0, \quad 0 < \lim_{p \to \infty} \frac{p}{m+n} < 1.
\]
satisfies $0 < d_1 < 1$ and, $0 < d_2 < \infty$.
Let,
\[
\breve{m} = \max\{m, p\}, \quad \breve{n} = \min\{n, m+n-p\}, \quad \breve{p} = \min\{m, p\}.
\]
Moreover, let
\[
\sin^2\left(\frac{\gamma}{2}\right) = \frac{\min\{\breve{p}, \breve{n}\} - \frac{1}{2}}{\breve{m} + \breve{n} - 1}, \quad \sin^2\left(\frac{\psi}{2}\right) = \frac{\max\{\breve{p}, \breve{n}\} - \frac{1}{2}}{\breve{m} + \breve{n} - 1},
\]
\begin{equation}
\mu_{J,p} = \tan^2 \left(\frac{\gamma + \psi}{2}\right),
\end{equation}
\begin{equation}
\sigma^3_{J,p} = \frac{16 \mu^3_{J,p}}{ (\breve{m} + \breve{n} - 1)^2} \cdot \frac{1}{\sin(\gamma)\sin(\psi)\sin^2(\gamma + \psi)}.
\end{equation}
Set 
\[
\mathbf{B}_p = \frac{\mathbf{X}\mathbf{X}^T }{\breve{n}} \quad \text{and} \quad \mathbf{A}_p = \frac{\mathbf{Y}\mathbf{Y}^T }{\breve{m}} .
\] Denote the largest root of
\[
\det(\lambda \mathbf{A}_p - \mathbf{B}_p) = 0
\]
by $\lambda_1$. Then
\[
\lim_{p \to \infty} P\left( \frac{\frac{\breve{n}}{\breve{m}} \lambda_1 - \mu_{J,p}}{\sigma_{J,p}} \leqslant s \right) = F_1(s)
\]
where $F_1(s)$ is the cumulative distribution function of the Tracy-Widom distribution defined in \Cref{eq:TW cdf}. 
\end{theorem}

The above theorem has a couple of interesting remarks. For instance, this immediately implies the distribution of the largest root of $\det(\lambda(\mathbf{B}_p + \mathbf{A}_p) - \mathbf{B}_p) = 0$. In fact, the largest root of $\det(\lambda(\mathbf{B}_p + \mathbf{A}_p) - \mathbf{B}_p) = 0$ is $\frac{\lambda_1}{1 + \lambda_1}$ if $\lambda_1$ is the largest root of the F matrices $\mathbf{B}_p \mathbf{A}_p^{-1}$ in \Cref{th: tw for Fmat} when $0 < d_1 < 1$.

When $d_1 > 1$, the largest root of $\det(\lambda(\mathbf{B}_p + \mathbf{A}_p) - \mathbf{B}_p) = 0$ is one with multiplicity $(p - m)$. In that case, instead one considers the $(p - m + 1)$th largest root of $\det(\lambda(\mathbf{B}_p + \mathbf{A}_p) - \mathbf{B}_p) = 0$. It turns out that the $(p - m + 1)$th largest root of $\det(\lambda(\mathbf{B}_p + \mathbf{A}_p) - \mathbf{B}_p) = 0$ is $\frac{\lambda_1}{1 + \lambda_1}$ if $\lambda_1$ is the largest root of $\det(\lambda \mathbf{A}_p - \mathbf{B}_p) = 0$. The exact order of the centering and scaling parameters $\mu_{J,p}$ and $\sigma_{J,p}$ can also be obtained along the lines of this result in terms of $\breve{m},\breve{n}$ and $p$. \Cref{sec:application} has a detailed discussion on the applications of these results on high-dimensional inference.

\section{Applications in Statistics}
\label{sec:application}
In this section, we discuss various applications of random matrix theory in statistics and related fields. So far there are a lot of ground-breaking applications of RMT that helped to develop robust, efficient high dimensional data handling methods, enriched complex machine learning algorithms, optimized signal processing techniques and motivated a lot of crucial discoveries in genomics, finance, climate science, and social network analysis. Henceforth, in this section, the key focus is on these applications to real-world problems in conjunction with the theoretical discussion above. 

\subsection{Inference on Covariance Matrices}
\label{sebsec: infcovmat}
One of the primary applications of the theory of large random matrices in high-dimensional statistics is inference on covariance matrices. Since the results provide asymptotic properties of the spectra of large random matrices, using those results one can check whether one estimate is consistent under certain conditions, and also for hypothesis testing, one can approximate the power functions if the test statistic is a function of the spectrum. In this regard, One of the earliest uses of the distribution of the largest eigenvalue of the sample covariance matrix is in testing the hypothesis \( H_0 : \Sigma = \mathbf{I}_p \) when i.i.d. samples are drawn from a \( N(\mu, \Sigma) \) distribution. The Tracy--Widom law for the largest sample eigenvalue under the null Wishart case, i.e., when the population covariance matrix \( \Sigma = \mathbf{I}_p \), allows a precise determination of the cut-off value for this test, which, with a careful calibration of the centering and normalizing sequences, is very accurate even for relatively small \( p \) and \( n \) (\citet{johnstone2001distribution},\citet{johnstone2009sparse}). 

The behavior of the power of the test requires formulating suitable alternative models. For instance, for data matrix $\mathbf{X} = [X_1,\cdots,X_p] \in \R^{p \times n}$ with iid columns $x_i$, consider the testing problem, 
\begin{equation}
    \begin{split}
        H_0&: \mathbf{X} = \sigma \mathbf{Z} \\
        H_1&: \mathbf{X} = a \mathbf{s}^T + \sigma \mathbf{Z}
    \end{split}
\end{equation}
where $\mathbf{Z} = [\mathbf{z}_1, \ldots, \mathbf{z}_n] \in \mathbb{R}^{p \times n}$ with $\mathbf{z}_i \sim N(\mathbf{0}, \mathbf{I}_p)$, $\mathbf{a} \in \mathbb{R}^p$ deterministic with unit norm $\|\mathbf{a}\| = 1$, $\mathbf{s} = [s_1, \ldots, s_n]^\top \in \mathbb{R}^n$ with $s_i$ i.i.d. random scalars, and $\sigma > 0$. We also denote $c = p / n$ (and demand as usual that $0 < \lim \inf c \leq \lim \sup c < \infty$).

This model describes the observation of either pure Gaussian noise data $\sigma \mathbf{z}_i$ with zero mean and covariance $\sigma^2 \mathbf{I}_p$, or of deterministic information $\mathbf{a}$ possibly modulated by a scalar (random) signal $s_i$ (which could simply be $\pm 1$) added to the noise. If the parameters $\mathbf{a}$, $\sigma$ as well as the statistics of $s_i$ are known, a mere Neyman-Pearson test allows one to discriminate between $H_0$ and $H_1$ with optimal detection probability, for all finite $n, p$; precisely, one will decide on the genuine hypothesis according to the ratio of posterior probabilities
\begin{equation}
    \label{eq: LRT}
    \frac{\mathbb{P}(\mathbf{X} \mid H_1)}{\mathbb{P}(\mathbf{X} \mid H_0)} 
\underset{H_0}{\overset{H_1}{\gtrless}} \alpha
\end{equation}
for some $\alpha > 0$ controlling the desired Type I and Type II error rates (that is, the probability of false positives and of false negatives).

However, in practice, unless the existence of a set of previous pure-noise acquisitions is assumed, it is quite unlikely that $\sigma$ be assumed known or consistently estimated. Similarly, if the ultimate objective (post-decision) is to estimate the data structure $\mathbf{a}$ under $H_1$, $\mathbf{a}$ is naturally assumed partially or completely unknown (it may be known to belong to a subset of $\mathbb{R}^p$ in which case more elaborate procedures than proposed here can be carried on). In the most generic scenario where $\mathbf{a}$ is fully unknown, assuming additionally the data of zero mean, we may thus impose without generality the restriction that Under this (very restricted) prior knowledge, instead of the maximum likelihood test in (\ref{eq: LRT}), one may resort to a \textit{generalized likelihood ratio test (GLRT)} defined as
\[
\frac{\sup_{\sigma, \mathbf{a}} \mathbb{P}(\mathbf{X} \mid \sigma, \mathbf{a}, \mathcal{H}_1)}{\sup_{\sigma, \mathbf{a}} \mathbb{P}(\mathbf{X} \mid \sigma, \mathbf{a}, \mathcal{H}_0)} 
\underset{\mathcal{H}_0}{\overset{\mathcal{H}_1}{\gtrless}} \alpha.
\]

Under both Gaussian noise and signal $s_i$ assumption, the GLRT has an explicit expression that appears to be a monotonously increasing function of $\|\mathbf{X}\mathbf{X}^\top\| / \mathrm{tr}(\mathbf{X}\mathbf{X}^\top)$. That is, the test is equivalent to
\[
T_p \equiv \frac{\|\frac{1}{n}\mathbf{X}\mathbf{X}^\top\|}{\frac{1}{p} \mathrm{tr} \left( \frac{1}{n}\mathbf{X}\mathbf{X}^\top \right)}
\underset{\mathcal{H}_0}{\overset{\mathcal{H}_1}{\gtrless}} f(\alpha),
\]
 (\citet{wax1985detection} and \citet{anderson1963asymptotic} has a detailed discussion on this idea) for some known monotonously increasing function $f$. Here we introduced the normalizations $1/p$ and $1/n$ so that both the numerator and denominator are of order $O(1)$ as $n, p \to \infty$.

Since the ratio $T_p$ has limit $(1+\sqrt{c})^2$ under the $H_0$ asymptotics, $f(\alpha)$ must be of the form $f(\alpha) = (1+\sqrt{c})^2 + g(\alpha)$ for some $g(\alpha) > 0$. Also, as we know that $\frac{1}{p} \mathrm{tr} \left( \frac{1}{n} \mathbf{X}\mathbf{X}^\top \right)$ fluctuates at the speed $O(n^{-1})$, while $\|\frac{1}{n}\mathbf{X}\mathbf{X}^\top\|$ fluctuates at the slower speed $O(n^{-2/3})$ (as per \Cref{th: tracy widom}), the global fluctuation is dominated by the numerator at a rate of order $O(n^{-2/3})$, i.e., we have under $H_0$,
\[
T_p \stackrel{H_0}{=} (1 + \sqrt{c})^2 + O(n^{-2/3}).
\]
Since the denominator essentially converges (at an \( O(n^{-1}) \) rate) while the numerator still fluctuates (at an \( O(n^{-2/3}) \) rate), despite the dependence between both, only the fluctuations of the numerator \( \frac{1}{n} \mathbf{X}\mathbf{X}^\top \) influence the behavior of the ratio \( T_p \), and thus
\[
T_p \overset{H_0}{\sim} (1 + \sqrt{c})^2 + (1 + \sqrt{c}) \frac{4}{3} c^{-\frac{1}{6}} n^{-\frac{2}{3}} \text{TW} + o(n^{-2/3}),
\]
where TW denotes the Tracy-Widom Distribution. As a consequence, in order to set a maximum false alarm rate (or false positive, or Type I error) of \( r > 0 \) in the limit of large \( n, p \), one must choose a threshold \( f(\alpha) \) for \( T_p \) such that
\[
\mathbb{P}(T_p \geq f(\alpha)) = r,
\]
that is, such that
\[
\mu_{\text{TW}}([A_p, +\infty)) = r, \quad A_p = (f(\alpha) - (1 + \sqrt{c})^2)(1 + \sqrt{c})^{-\frac{4}{3}} c^{\frac{1}{6}} n^{\frac{2}{3}} \tag{3.2}
\]
with \( \mu_{\text{TW}} \), the Tracy-Widom measure.

For testing problems on covariance matrices, with a both-sided alternative, based on a normal random sample, instead of Tracy-Widom law, one can also use \Cref{th:th1}. For $X_1,\cdots,X_n \overset{iid}{\sim} N_p(\mu,\mathbf{\Sigma})$ where $\mu,\Sigma$ are unknown and $p,n$ are both large and, $p \sim n^{\frac{1}{2} - \epsilon}$, $\epsilon > 0$. Consider the testing problem with a two-sided alternative,
\begin{equation}
\label{eq: two sided test}
    \begin{split}
        H_0: \mathbf{\Sigma} = \mathbf{\Sigma}_0 \\
        H_1: \mathbf{\Sigma} \neq \mathbf{\Sigma}_0
    \end{split}
\end{equation}
From \Cref{th:th1}, it turns out 
\begin{equation}
    \phi(\mathbf{X}) = \mathbf{1}\left( \sqrt{\frac{n-1}{2p}} \left| \sum_{i=1}^p \log\left( \frac{\hat{\lambda}_i}{\lambda_i} \right) - \sum_{i=1}^p \log(n-p+i) 
  \right| > z_{\alpha/2}  \right)
\end{equation}
is an asymptotically size $\alpha$ test, where $\hat{\lambda}_1,\cdots,\hat{\lambda}_p$ are the eigenvalues of the sample covariance matrix $S = \frac{1}{n-1} \sum_{i=1}^n (X_i - \overline{X}_n)(X_i - \overline{X}_n)^T$, $\lambda_1,\cdots,\lambda_p$ are the eigenvalues of $\mathbf{\Sigma}_0$, $\alpha \in (0,1)$ and $z_{\alpha}$ is the $(1-\alpha)$th quantile of N(0,1). 

The same idea can be generalized for testing problems of the covariance matrices in high-dimensional Linear Regression when the number of response variables and number of data points are both large. Consider the multivariate Linear Regression model, 
\begin{equation}
\label{eq: High dim Lin Reg}
    \mathbf{Y} = \mathbf{XB} + \mathbf{E}
\end{equation}
where $\mathbf{Y} = [Y_1: \cdots: Y_m] \in \R^{n \times m}$ is the response matrix consisting $n$ observations for each of the $m$ response variable, $\mathbf{X} \in \R^{n \times p}$ is the design matrix where $p$ is the number of covariates. $\mathbf{E}$ denotes the error matrix. For inference purposes, it is further assumed that $\mathbf{E} \sim NDM(0,\mathbf{\Sigma})$ i.e. rows of $\mathbf{E}$ are iid from $N(0,\mathbf{\Sigma})$. Consider the testing problem with a two-sided alternative, as in (\ref{eq: two sided test}) i.e. 
\begin{equation*}
    \begin{split}
        \begin{split}
        H_0: \mathbf{\Sigma} = \mathbf{\Sigma}_0 \\
        H_1: \mathbf{\Sigma} \neq \mathbf{\Sigma}_0
    \end{split}
    \end{split}
\end{equation*}
Under $H_0$, the sum of squares of error (SSE) defined as $\mathbf{Y}^T(\mathbf{I - P_X})\mathbf{Y}$, where $\mathbf{P_X}$ is the orthogonal projection matrix of $\mathcal{C}(\mathbf{X})$ follows $W_m(\mathbf{\Sigma}_0,n-r)$ with $r$ to be the rank of $\mathbf{X}$. Therefore an asymptotically level $\alpha$ test to test (\ref{eq: two sided test}) is given by 
\begin{equation}
    \phi_{\mathcal{R}} := \mathbf{1}\left( \sqrt{\frac{n-r}{2m}} \left| \sum_{i=1}^m \log\left( \frac{\hat{\lambda}_i}{\lambda_i} \right) - \sum_{i=1}^m \log(n-r-k+i) 
  \right| > z_{\alpha/2}  \right)
\end{equation}

where $\hat{\lambda}_1,\cdots,\hat{\lambda}_p$ are the eigenvalues of SSE $= \mathbf{Y}^T(\mathbf{I - P_X})\mathbf{Y}$, $\lambda_1,\cdots,\lambda_p$ are the eigenvalues of $\mathbf{\Sigma}_0$, $\alpha \in (0,1)$ and $z_{\alpha}$ is the $(1-\alpha)$th quantile of N(0,1). 

The same idea can be generalized for the two sample tests of equality for covariance matrices as well. Suppose we have data points $X_1,\cdots,X_m \overset{iid}{\sim} N_p(\mu_1,\mathbf{\Sigma}_1)$ and $Y_1,\cdots,Y_n \overset{iid}{\sim} N_p(\mu_2,\mathbf{\Sigma}_2)$ where $\mu_1,\mu_2,\mathbf{\Sigma}_1,\mathbf{\Sigma}_2$ are unknown. $n \equiv n(m)$ satisfies $\lim_{m \to \infty} \frac{n}{m} = c \in (0,\infty)$. Consider the testing problem, 
\begin{equation}
    \begin{split}
        H_0: \mathbf{\Sigma_1 = \Sigma_2} \\
        H_1: \mathbf{\Sigma_1 \neq \Sigma_2}
    \end{split}
\end{equation}
Then by \Cref{th:th1} it turns out that
\begin{equation}
    \begin{split}
        \phi(\mathbf{X,Y}) := \mathbf{1}\left( \sqrt{\frac{m}{2p\left( 1 + \frac{1}{c} \right)}} \left| \sum_{i=1}^p \log\left( \frac{\hat{\lambda}_i}{\hat{\lambda}_i^*} \right) - \sum_{i=1}^p \log \left( \frac{n-p+i}{m-p+i} \right) 
  \right| > z_{\alpha/2}  \right)
    \end{split}
\end{equation}
is an asymptotically level $\alpha$ test where $\hat{\lambda}_1,\cdots,\hat{\lambda}_p$ are the eigenvalues of the sample covariance matrix $\mathbf{S_x} = \frac{1}{m-1} \sum_{i=1}^m (X_i - \overline{X}_m)(X_i - \overline{X}_m)^T$, $\hat{\lambda}_1^*,\cdots,\hat{\lambda}_p^*$ are the eigenvalues of $\mathbf{S_Y} = \frac{1}{n-1} \sum_{i=1}^n (Y_i - \overline{Y}_n)(Y_i - \overline{Y}_n)^T$, $\alpha \in (0,1)$ and $z_{\alpha}$ is the $(1-\alpha)$th quantile of N(0,1). The power function for this test is given by, 
$\beta(\mathbf{\Sigma_1,\Sigma_2}) := 1 - \Phi\left( z_{\alpha/2} -  \sqrt{\frac{m}{2p\left( 1 + \frac{1}{c} \right)}}\sum_{i=1}^p \log\left(\frac{\lambda_i}{\lambda_i^*} \right)  \right) + \Phi\left( - z_{\alpha/2} -  \sqrt{\frac{m}{2p\left( 1 + \frac{1}{c} \right)}}\sum_{i=1}^p \log\left(\frac{\lambda_i}{\lambda_i^*} \right)  \right) + f(n,m)$
where $f(n,m) = O\left( p\left( \frac{1}{\sqrt{m}} + \frac{1}{\sqrt{n}} \right) \right)$.

The same test can be done using the asymptotic theory of the largest root of $F-$type matrices as well. Since $\mathbf{S_Y}$ is almost surely invertible, we can define the test, 
\begin{equation}
    \phi_{\mathcal{F}} := \mathbf{1}\left( \frac{\frac{\breve{n}}{\breve{m}} \lambda_1 - \mu_{J,p}}{\sigma_{J,p}} > F_1^{-1}(1-\alpha) \right)
\end{equation}
where $\breve{n},\breve{m},\mu_{J,p},\sigma_{J,p},F_1(\cdot)$ are as in \Cref{th: tw for Fmat} and $\lambda_1$ is the largest root of
\[\big((n-1)\mathbf{S_Y}\big)^{-1}\big((m-1)\mathbf{S_X} \big)\]
 By \Cref{th: tw for Fmat}, $\phi_{\mathcal{F}}$ turns out to be an asymptotic size $\alpha$ test. 

 For the high-dimensional linear regression model defined in (\ref{eq: High dim Lin Reg}), one can also obtain a high-dimensional generalization of Wald's test using the asymptotic theories of the $F-$type matrices. Consider the wald's testing problem 
 \begin{equation}
     \begin{split}
         H_0: \mathbf{L^TB = B_0}\\
         H_1: \mathbf{L^TB \neq B_0}
     \end{split}
 \end{equation}
 where $\mathbf{L} \in \R^{p \times k}$ and rank of $\mathbf{L}$ is $k$. If the design matrix $\mathbf{X}$ has full column rank, i.e. $\rho(\mathbf{X}) = p$, the Ordinary Least Squares (OLS) estimate for $\mathbf{B}$ is given by $\mathbf{\hat{B}} = \mathbf{(X^TX)^{-1}X^TY}$. Then under $H_0$, 
 \begin{equation}
     \mathbf{A_p} := \mathbf{(L^T\hat{B} - B_0)^T (L^T(X^TX)^{-1}L)^{-1}(L^T\hat{B} - B_0)} \overset{H_o}{\sim} W_m(\mathbf{\Sigma},k)
 \end{equation}
 and 
 \begin{equation}
     \mathbf{B_p = Y^T(I-P_X)Y} \sim W_m(\mathbf{\Sigma},n-p)
 \end{equation}
 and $\mathbf{A_p,B_p}$ are independent. Let $\lambda_1$ be the largest root of $det(\lambda A_p - B_p)$. Then from \Cref{th: tw for Fmat}, it turns out that a normalized version of $\lambda_1$ asymptotically follows Tracy-Widom Distribution under $H_0$. Therefore, in view of \Cref{th: tw for Fmat}, one can construct an asymptotic size $\alpha$ test using $\lambda_1$.

\subsection{Application in PCA}
\label{subsec: app in PCA} 
In \Cref{subsec:cov matrices} we discussed the LSD and asymptotic properties of the sample covariance matrix under the Gaussianity assumption when the eigenvalues of the population covariance matrix are either identical or are evenly spread out so that none of them “sticks out” from the bulk. \citet{soshnikov2002note} proved the
 distributional limits under weaker assumptions, in addition to deriving distributional limits of the $k-$th largest eigenvalue, for fixed but arbitrary $k$. Under the Gaussianity assumption of the data, the asymptotic distribution of the eigenvalues of the sample covariance matrix also turns out to be Gaussian if the eigenvalues of the population covariance matrix are distinct. 

 \begin{theorem}[\citet{mardia2024multivariate}]
     Let $X_1,\cdots,X_n \overset{iid}{\sim} N_p(0,\mathbf{\Sigma})$ where $\mathbf{\Sigma}$ is positive definite with all distinct eigenvalues $\lambda_1 > \cdots \lambda_p > 0$. Let $l_{n,1} \geqslant \cdots \geqslant l_{n,p}$ be the eigenvalues of $\frac{1}{n} \sum_{i=1}^nX_iX_i^T$ and $\mathbf{\Lambda} = \text{diag}(\lambda_1,\cdots,\lambda_p)$. Then 
     \begin{equation}
         \sqrt{n}(l_n - \mathbf{\lambda}) \convD N_p(0,2\mathbf{\Lambda}^2)
     \end{equation}
     as $n \to \infty$ where $l_n = \begin{bmatrix}
         l_{n,1}\\
         \vdots \\
         l_{n,p}
     \end{bmatrix}$ and $\lambda = \begin{bmatrix}
         \lambda_1\\
         \vdots \\
         \lambda_p
     \end{bmatrix}$
 \end{theorem}

So the consistency of the sample eigenvalues of the sample covariance matrix holds when the population covariance matrix has either all eigenvalues identical or all distinct from each other. However, in recent years, researchers in various fields have been using different versions of covariance matrices of growing dimensions with special patterns. For instance,  in speech recognition (\citet{hastie1995penalized}),  wireless communication (\citet{telatar1999capacity}), and statistical learning (\citet{hoyle2003limiting}) a few of the sample eigenvalues have limiting behavior that is different from the behavior when the covariance is the identity.

While high-dimensional data often exhibits complex patterns, it's frequently characterized by a simple underlying structure. This structure can be modeled as a low-dimensional "signal" obscured by high-dimensional "noise." Assuming an additive relationship between these components, we can represent the data using a factor model. Factor models are particularly useful for detecting and estimating low-dimensional signals within isotropic or nearly isotropic noise. Key statistical questions, such as those related to dimension reduction, can be effectively addressed by analyzing the eigenvalues and eigenvectors of the sample covariance matrix. A particularly useful
 idealized model of this kind, named the spiked covariance model by \citet{johnstone2001distribution} has been in use for quite some time in statistics. Under this model, the population covariance matrix $\mathbf{\Sigma}$ is expressed as
 
 \begin{equation}
 \label{eq: spike}
\mathbf{\Sigma} = \sum_{j=1}^M \lambda_j \theta_j \theta_j^* + \sigma^2 I_p,
\end{equation}
where $\theta_1, ..., \theta_M$ are orthonormal; $\lambda_1 \geq ... \geq \lambda_M > 0$ and $\sigma^2 > 0$. This model implies that, except for $M$ leading eigenvalues $l_j = \lambda_j + \sigma^2$ for $j = 1, ..., M$, the rest of the eigenvalues are all equal.

This model has been studied extensively in the context of high-dimensional PCA since it brings out several key issues associated with dimension reduction in the high-dimensional context. \citet{johnstone2009sparse} first demonstrated that if $\frac{p}{n} \to \gamma \in (0,\infty)$ the sample principal components are inconsistent estimates of the population principal components under (\ref{eq: spike}). This phase transition phenomenon is described in its simplest form in the following theorem, where, for convenience, we assume in (\ref{eq: spike}) $\sigma^2 = 1$.

\begin{theorem}[\citet{baik2006eigenvalues}]
\label{th:th 10}
    Suppose that $\Sigma$ is a $p \times p$ positive definite matrix with eigenvalues $\ell_1 \geq \cdots \geq \ell_M > 1 = \cdots = 1$, and let $\hat{\ell}_1 \geq \cdots \geq \hat{\ell}_p$ be the eigenvalues of the sample covariance matrix $S = n^{-1} \Sigma^{1/2} Z Z^* \Sigma^{1/2}$ where the $p \times n$ data matrix $Z$ has i.i.d. real or complex entries with zero mean, unit variance and finite fourth moment. Suppose that $p, n \to \infty$ such that $p/n \to \gamma \in (0, \infty)$. Then, for each fixed $j=1,2,\cdots,M$
    \begin{equation}
    \hat{\ell}_j \xrightarrow{\text{a.s.}} \begin{cases}
    (1 + \sqrt{\gamma})^2 & \text{if } \ell_j \leq 1 + \sqrt{\gamma}, \\
    \ell_j \left(1 + \frac{\gamma}{\ell_j - 1}\right) & \text{if } \ell_j > 1 + \sqrt{\gamma}.
\end{cases}
\end{equation}

\end{theorem}

Therefore, when the population covariance matrix is of the spike form, it might not be such a good idea to use Principal Component Analysis (PCA) for dimension reduction in a high-dimensional setting, at least not in its standard form. In this regard, one natural question is how one can test if the population covariance matrix $\mathbf{\Sigma}$ is in the form of (\ref{eq: spike}) and how bad the inconsistency of the sample principal components if $\mathbf{\Sigma}$ is in spike form. The following theorem provides an answer to these questions. 

\begin{theorem}[\citet{paul2007asymptotics}]
\label{th: th 11}
    Suppose that $X_1,\cdots,X_n \overset{iid}{\sim} N_p(0,\Sigma)$ where $\Sigma$ is a $p \times p$ positive definite matrix with eigenvalues $\ell_1 \geq \cdots \geq \ell_M > 1 = \cdots = 1$, and let $\hat{\ell}_1 \geq \cdots \geq \hat{\ell}_p$ be the eigenvalues of the sample covariance matrix $S = \frac{1}{n}\sum_{i=1}^nX_iX_i^T$. Suppose that $p, n \to \infty$ such that $\frac{p}{n} - \gamma = o(n^{-1/2})$ for a $\gamma \in (0,\infty)$. For a fixed $j \in \{1,2,\cdots,M\}$ if $\ell_j > 1 + \sqrt{\gamma}$, then 
    \begin{equation}
        \sqrt{n}\left(\hat{\ell}_j - \ell_j \left(1 + \frac{\gamma}{\ell_j - 1}\right)  \right) \convD N(0,\sigma^2(\ell_j))
    \end{equation}
    as $n \to \infty$ where $\sigma^2(\ell) := 2\ell^2\left(1 - \frac{\gamma}{(\ell-1)^2} \right)$
    
\end{theorem}

Suppose that we test the hypothesis $H_0: \Sigma = I$ versus the alternative that $H_1: \Sigma =$ diag$(\ell_1, \ldots, \ell_M, 1, \ldots, 1)$ with $\ell_1 \geq \cdots \geq \ell_M > 1$, based on i.i.d. observations from $N(0, \Sigma)$. If $\ell_1 > 1 + \sqrt{\gamma}$, it follows from \Cref{th:th 10} that the largest root test is asymptotically consistent. For the special case when $\ell_1$ is of multiplicity one, \Cref{th: th 11} gives an expression for the asymptotic power function, assuming that $p / n$ converges to $\gamma$ fast enough, as $n \to \infty$. One has to view this in context since the result is derived under the assumption that $\ell_1, \ldots, \ell_M$ are all fixed, and we do not have a rate of convergence for the distribution of $\hat{\ell}_1$ toward normality. However, \Cref{th: th 11} can be used to find confidence intervals for the larger eigenvalues under the non-null model. 

Under the same set of assumptions as of \Cref{th: th 11}, \citet{paul2007asymptotics} proved further that, if $\ell_j \leqslant 1 + \sqrt{\gamma}$ and $\ell_j$ is of arithmetic multiplicity one, then the angle between the $j-$th sample and population eigenvectors converges to $\frac{\pi}{2}$ almost surely which essentially shows in which extent the sample principal components can be inconsistent and provides a generalization of \citet{johnstone2009sparse}. Later on \citet{bai2008large} extends the results of \citet{paul2007asymptotics} in the context of spiked covariance matrix by dropping the Gaussianity assumption. 
\begin{figure}[h]
    \centering
    \includegraphics[width=0.7\linewidth]{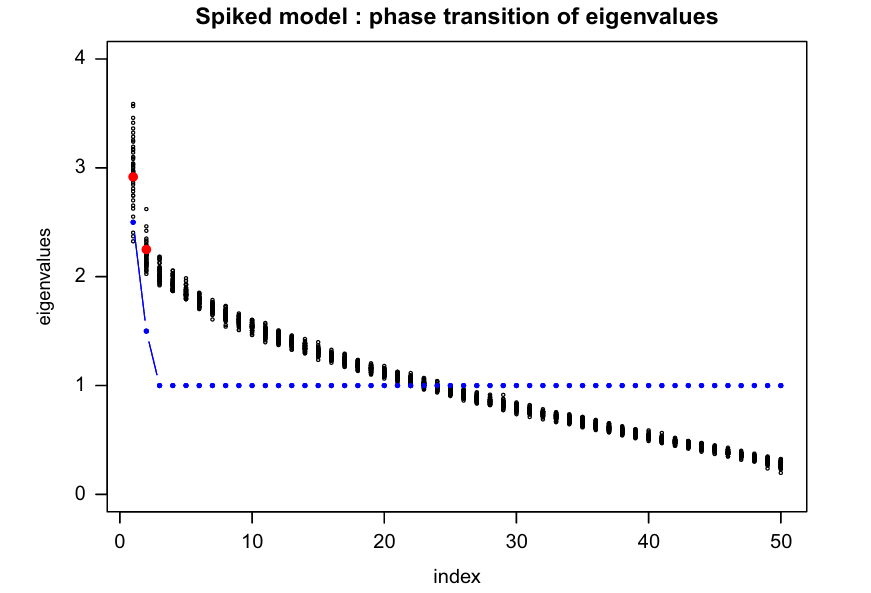}
    \caption{(\citet{paul2007asymptotics}) An illustration of the phase transition of eigenvalues in a spiked covariance model: here, $p=50$, $n=200$ and eigenvalues of the covariance matrix are $\ell_1=2.5$, $\ell_2=1.5$, $\ell_j=1$ for $j=3,\ldots,p$. So, $\ell_1 > 1+\sqrt{p/n}$ and $\ell_2 = 1+\sqrt{p/n}$. Blue dots correspond to the population eigenvalues. Black circles correspond to the sample eigenvalues (based on i.i.d. Gaussian samples) for 50 replicates. Solid red circles indicate the theoretical limits of the first two eigenvalues for $\gamma=p/n=0.25$.}
    \label{fig:enter-label}
\end{figure}

\subsection{On signal processing and wireless communications}
\label{subsec: signal pro}

Large random matrices come up often in signal processing, especially in wireless communication. \citet{bai2010spectral}, \citet{couillet2011random}, and \citet{tulino2004random} highlight several such cases, including: (i) finding the channel capacity of MIMO (multiple-input-multiple-output) systems, which involves calculating the logarithm of the determinant of the matrix \( \mathbf{I + S} \), where \( \mathbf{S} \) is a Wishart matrix that reflects the signal-to-noise ratio in transmission; (ii) finding the limiting SINR (signal-to-interference-noise ratio) in random channels and in linearly precoded systems, like CDMA (code-division-multiple-access) systems (\citet{bai2007signal}); (iii) analyzing the performance of receivers as the system size grows; and (iv) estimating energy from multiple sources (\citet{couillet2011random}). Besides, random matrices are useful in a variety of signal-processing problems, such as detecting input signals (\citet{nadakuditi2010fundamental}; \citet{silverstein1992signal}) and estimating subspaces in sensor networks (\citet{hachem2013subspace}). Furthermore, the asymptotic distribution of the spectra of large random matrices and the idea behind Roy's largest root test (\citet{roy1953heuristic}, \citet{johnstone2017roy}) can be used to construct nonparametric tests to detect the number of signals embedded in noise (\citet{kritchman2009non}).  

The standard setup for signals impinging on an array with sensors consists of $n$ i.i.d $p-$dimensional observations $\{\mathbf{x}_i\}_{i=1}^n$ from the model, 
\begin{equation}
    \mathbf{x}(t) = \mathbf{A} \mathbf{s}(t) + \sigma \mathbf{n}(t)
\end{equation}
sampled at $n$ distinct times $t_i$, where $\mathbf{A} = [\mathbf{a}_1, \dots, \mathbf{a}_K]$ is the $p \times K$ steering matrix of $K$ linearly independent $p$-dimensional vectors. The $K \times 1$ vector $\mathbf{s}(t) = [s_1(t), \dots, s_K(t)]^T$ represents the random signals, assumed zero mean and stationary with full rank covariance matrix. $\sigma$ is the unknown noise level, and $\mathbf{n}(t)$ is a $p \times 1$ additive Gaussian noise vector, distributed $\mathcal{N}(0, \mathbf{I}_p)$ and independent of $\mathbf{s}(t)$.

Under these assumptions, the population covariance matrix $\Sigma$ of $\mathbf{x}(t)$ has a diagonal form,
\begin{equation}
    \mathbf{W}^H \Sigma \mathbf{W} = \sigma^2 \mathbf{I}_p + \text{diag}(\lambda_1, \dots, \lambda_K, 0, \dots, 0)
\end{equation}
where columns of $\mathbf{W}$ forms a basis of $\mathbb{C}^p$ (or of $\mathbb{R}^p$ if the signals are real valued). Let $\mathbf{S}_n$ be the sample covariance matrix of $\{\mathbf{x}_i\}_{i=1}^n$, defined as 
\begin{equation*}
    \mathbf{S}_n = \frac{1}{n} \sum_{i=1}^n \mathbf{x}_i\mathbf{x}_i^H
\end{equation*}
having the eigenvalues $l_1 \geqslant l_2 \geqslant \cdots \geqslant l_p$.  

The number of signals $K$, can then be estimated with the number of eigenvalues of the sample covariance matrix $\mathbf{S}_n$ which are \textit{significanly larger} i.e. bigger than a certain threshold, where the individual thresholds for the eigenvalues can be determined using the Tracy Widom laws (\Cref{th: tracy widom}).  The following algorithm, which is deeply motivated by Roy's largest root test (\citet{roy1953heuristic}, \citet{johnstone2017roy}), takes the eigenvalues $l_1,\cdots,l_p$ of the sample covariance matrix $\mathbf{S}_n$ as input and gives the estimated number of signals $\hat{K}_{\text{RMT}}$ as output. The algorithm works as follows: For $k = 1, \dots, \min(p, n) - 1$, we test
\[
H_0 : \text{at most } k - 1 \text{ signals} \quad \text{vs.} \quad H_1 : \text{at least } k \text{ signals}.
\]
Under the null hypothesis, $\ell_k$ arises from noise. Thus, we reject $H_0$ if $\ell_k$ is too large, i.e. 
\[
\ell_k > \hat{\sigma}^2(k) C_{n,p,k}(\alpha)
\]
where $\hat{\sigma}^2(k)$ is an estimate for the unknown noise level $\sigma^2$ taken to be, 
\begin{equation}
\label{eq: noise est}
    \hat{\sigma}^2(k) = \frac{1}{p-k} \sum_{j = k+1}^p l_j
\end{equation}
 and 
 \begin{equation}
 \label{eq: cut off}
     C_{n,p,k}(\alpha) = \mu_{n,p - k} + s(\alpha) \xi_{n,p - k}
 \end{equation}
 where $\mu_{n,p}$ and $\xi_{n,p}$ are the centering and scaling parameters defined as
 \begin{equation}
    \begin{split}
        \mu_{n,p} &= \frac{1}{n} (\sqrt{n-1/2} + \sqrt{p-1/2})^2 \\
\xi_{n,p} &= \sqrt{\frac{\mu_{n,p}}{n}} \left(\frac{1}{\sqrt{n-1/2}} + \frac{1}{\sqrt{p-1/2}}\right)^{1/3}
    \end{split}
\end{equation}
and $s(\alpha)$ is the $1-\alpha$ quantile of the Tracy Widom distribution. \citet{kritchman2009non} showed, 
\[
\Pr\{\text{reject } H_0|H_0\} = \Pr\{\ell_k > \sigma^2 C_{n,p,k}(\alpha)|H_0\} \approx \alpha.
\]
Hence, $\alpha$ controls the probability of model overestimation. We stop at the smallest index $k$ where the above condition fails, i.e., the first time we accept $H_0$. Our estimate of the number of signals is then $\hat{K}_{\text{RMT}} = k - 1$. Hence, the estimator of the number of signals is, 
\[
\hat{K}_{\text{RMT}} = \arg \min_k \left\{ \ell_k < \hat{\sigma}^2(k)(\mu_{n,p - k} + s(\alpha) \xi_{n,p - k}) \right\} - 1.
\]

\begin{algorithm}[H]
\caption{Algorithm for detecting number of signals (\citet{kritchman2009non})}
\KwIn{Confidence level $\alpha$, observations $\ell_k$ for $k = 1, \dots, \min(p, n) - 1$}
\KwOut{Estimated number of signals $\hat{K}_{\text{RMT}}$}

\For{$k = 1$ \textbf{to} $\min(p, n) - 1$}{
    Compute the threshold $\hat{\sigma}^2(k) C_{n,p,k}(\alpha)$ using \ref{eq: noise est}, \ref{eq: cut off}\;
    \eIf{$\ell_k > \hat{\sigma}^2(k) C_{n,p,k}(\alpha)$}{
        conclude that there are at least $k$ signals and set $k = k+1$ \;
    }{
        conclude that there are at most $k-1$ signals\;
        Set $\hat{K}_{\text{RMT}} = k - 1$\;
        \textbf{break}\;
    }
}

\Return $\hat{K}_{\text{RMT}} = \arg \min_k \left\{ \ell_k < \hat{\sigma}^2(k)(\mu_{n,p - k} + s(\alpha) \xi_{n,p - k}) \right\} - 1$\;
\end{algorithm}

For a suitably chosen sequence of $\{\alpha\}_n$, $\hat{K}_{RMT,n}$ can be shown to be consistent i.e. $\underset{n \to \infty}{\text{lim}} \mathbb{P}(\hat{K}_{RMT,n} = K) = 1$ (\citet{kritchman2009non}) where $K$ is the original number of signals. 

To demonstrate the performance of the above algorithm, We plot the number of estimated signals when the actual number of signals is in a range of $2$ to $5$ and the errors are from Standard Normal, $t-$distribution with 5 df, Cauchy and Laplace distribution. Also, we vary the sample size $n$ in a range up to 5000. From \Cref{fig:est sig} it can be seen that, except when the noise has standard Cauchy distribution, if the sample size exceeds 1000, the estimated number of signals is the same as the original number of signals. The algorithm overestimates the number of signals for noise arriving from the Cauchy distribution. 

\begin{figure}[H]
    \centering
    \includegraphics[width=0.4\linewidth]{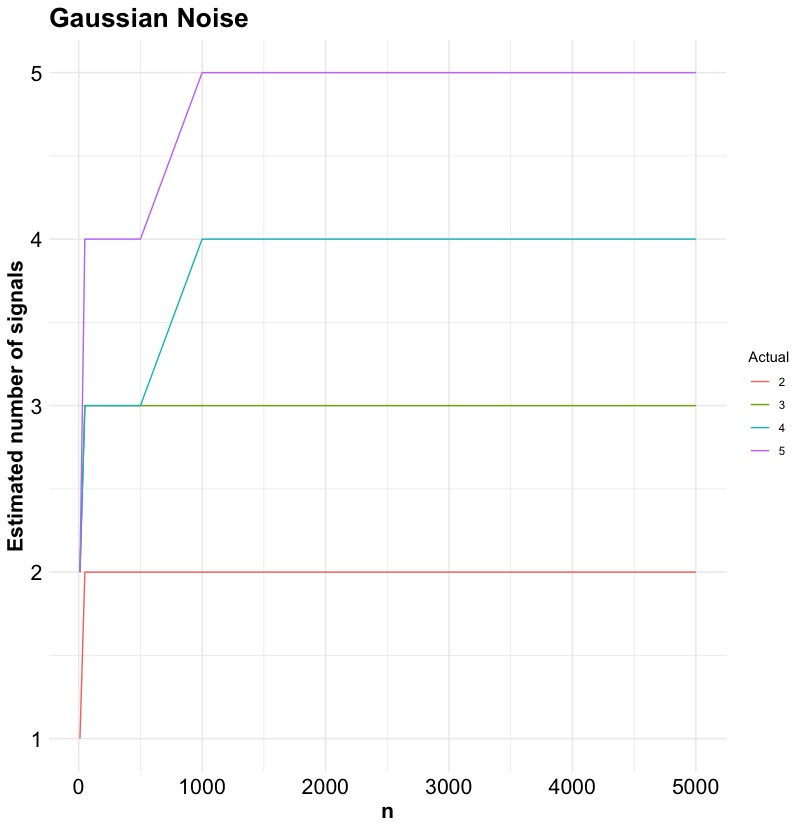}
    \includegraphics[width=0.4\linewidth]{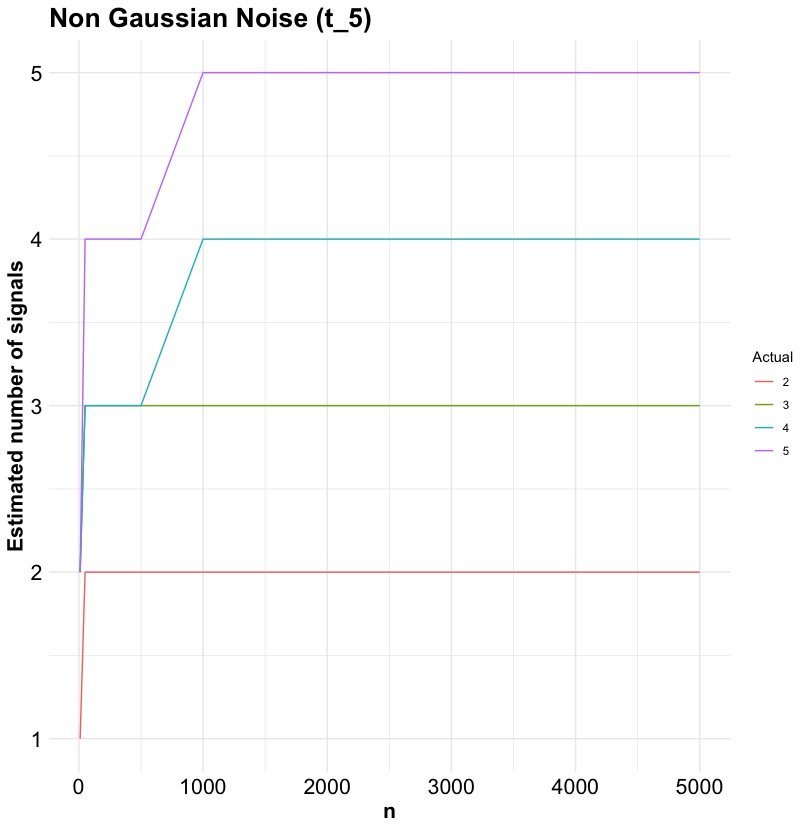}
    \includegraphics[width=0.4\linewidth]{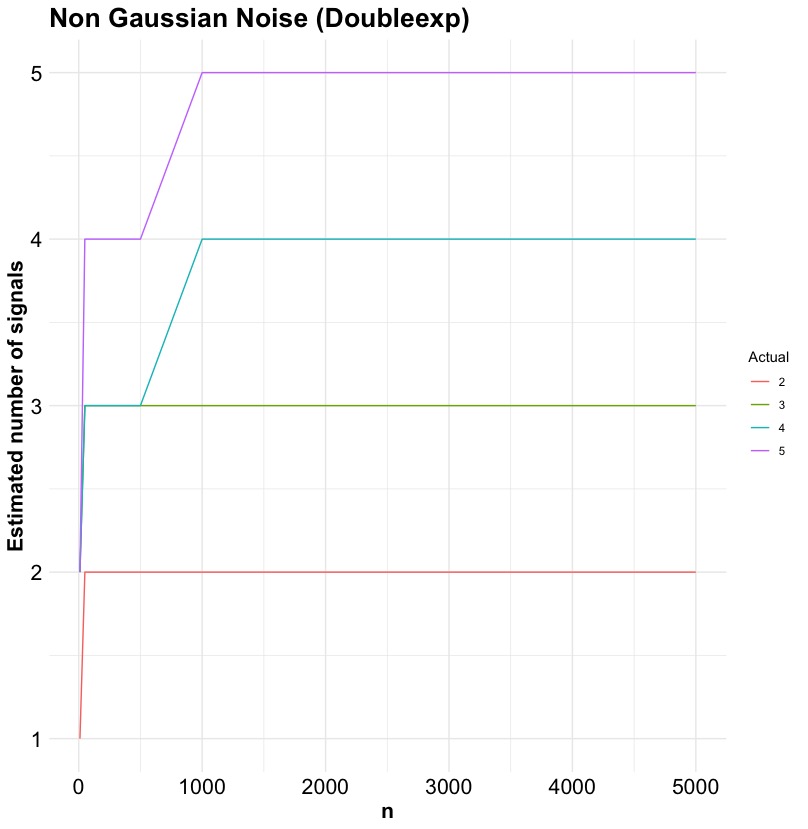}
    \includegraphics[width=0.4\linewidth]{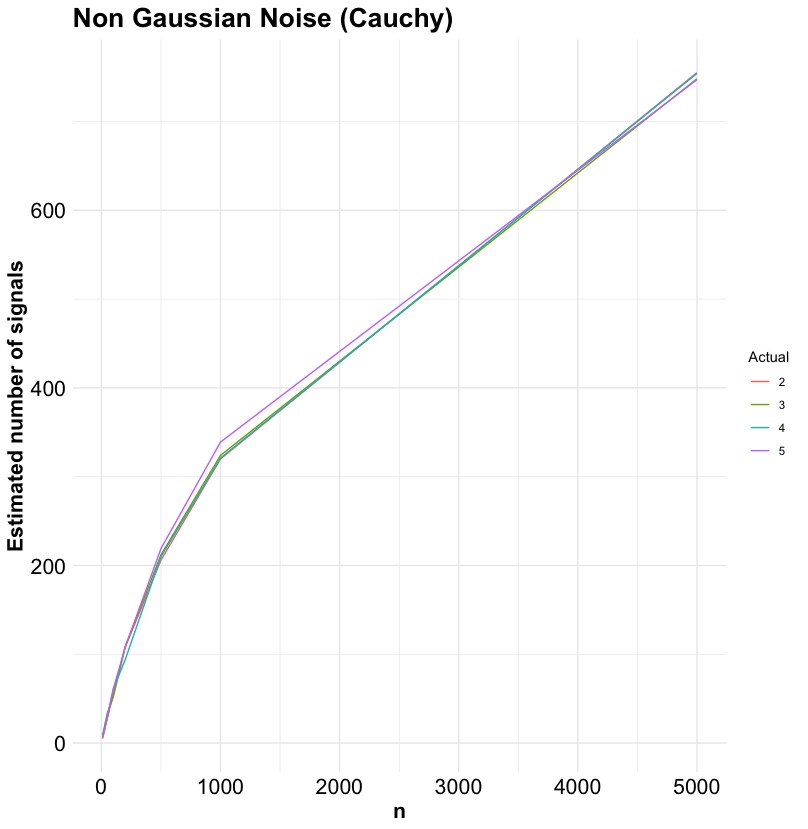}
    \caption{Number of Estimated Signals for different noise distributions: (in clockwise order) Standard Normal, t distribution with 5 df, Standard Cauchy, Double Exponential. The x-axis represents the sample size. }
    \label{fig:est sig}
\end{figure}


\subsection{On Changepoint Detection}
\label{subsec: change point, cov}

Change point detection (CPD) is a statistical method used to identify points in a dataset where the distribution of the data changes significantly. Studies of change-point detection problems date back to 1950. Since then, this topic has been of interest to statisticians and researchers in many other fields such as engineering,
 economics, climatology, biosciences, genomics, and
 linguistics due to its diverse applications. Different methods in parametric and nonparametric setups have been discovered (\citet{niu2016multiple}, \citet{aminikhanghahi2017survey}) for univariate and multivariate time series data. However, when the data is ultrahigh dimensional, most of these traditional methods struggle due to computational complexity or the failure to meet the underlying distributional assumptions. For example, to determine the change of covariance in high dimensional time series data, the sample covariance matrices are used which are extremely large dimensional. In this section, we discuss some methods of detecting change in covariance pattern of high dimensional time series which are motivated by the results of random matrix theory. Change point detection algorithms are traditionally classified as “online” or “offline.” We focus on the Offline setting, which considers the entire data set at once and looks back in time to recognize where the change occurred. 

    The literature on detecting changes in covariance for high dimensional time series has grown substantially in the last few years. Based on the theory of large-scale random matrices in \citet{hero2012hub}, \citet{banerjee2015non} has developed a method for covariance CPD when the data points are independently drawn from an unknown elliptically contoured distribution. \citet{avanesov2018change}, \citet{wang2017optimal} obtain method based on the distance between sample covariance matrices, using the operator norm and $l_\infty$ norm of matrices, respectively.

In particular, many authors consider changes in the moderate dimensional setting, that is, where the number of the parameters of the
model is of the order of the number of data points. \citet{ryan2023detecting} proposes a novel method for detecting changes in the covariance structure of moderate dimensional time series. Let \( X_1, \dots, X_n \in \mathbb{R}^p \) be independent \( p \)-dimensional vectors with
\begin{equation*}
\operatorname{cov}(X_i) = \Sigma_{i,p}, \quad \text{for } 1 \leq i \leq n,
\end{equation*}
where each \( \Sigma_{i,p} \in \mathbb{R}^{p \times p} \) is of full rank. Furthermore, let \( \mathbf{X}_{n,p} \) denote an \( n \times p \) matrix defined by \( \mathbf{X}_{n,p} := (X_1^T, \dots, X_n^T)^T \). The method primarily aims to develop a testing procedure that can identify a change in the covariance structure of the data over time. For now, let us consider the case of a single changepoint. We compare a null hypothesis of the data sharing the same covariance versus an alternative setting that allows a
 single change at time $\tau$. Formally we have
 \begin{equation}
     \begin{split}
         \label{eq:hypo}
H_0 &: \Sigma_{1,p} = \dots = \Sigma_{n,p} \\
H_1 &: \Sigma_{1,p} = \dots = \Sigma_{\tau,p} \neq \Sigma_{\tau+1,p} = \dots \Sigma_{n,p}
     \end{split}
 \end{equation}
where $\tau$ is unknown. We are interested in distinguishing between the null and alternative hypothesis, and under the alternative locating the changepoint $\tau$, when the dimension of the data $p$, is of comparable to the sample size, $n$. In particular, we require that for all pairs $n, p$, the set
\[T_{n,p}(\ell):=\{t\in \mathbb{Z}^+ \text{ such that } \ell < t < n-\ell\} \quad (2.4)\]
is nonempty, where $\ell > p$ is a problem dependent positive constant. Note $T_{n,p}(\ell)$ defines the set of possible candidate changepoints, while $\ell$ is the minimum distance between changepoints or minimum segment length. Then for each candidate changepoint $t\in T_{n,p}(\ell)$, a two-sample test statistic $T(t)$ can be used to determine if the data to the left and right of the changepoint have different distributions. If the two sample test statistic for a candidate exceeds some threshold, then we say a change has occurred and an estimator for $\tau$ is given by the value $t\in T_{n,p}(\ell)$ that maximizes $T(t)$.

In their method \citet{ryan2023detecting}, constructs the two sample test statistics using the eigenvalues of the sample covariance matrices of the two samples as follows. For two sample covariance matrices $\mathbf{A,B}$, in this context, we need to test whether $\mathbf{A}$ and $\mathbf{B}$ are equal or not. So in case they are identical, all of the eigenvalues of $R(\mathbf{A,B}) := \mathbf{B}^{-1}\mathbf{A}$ is 1. Therefore, the following function of the ratio matrix  (or $F-$type matrix) $R(\mathbf{A,B})$, gives a suitable measure of deviance from the equality of the two matrices, 
\begin{equation}
\label{eqn:Tstat}
    T(\mathbf{A}, \mathbf{B}) = \sum_{j=1}^{p} \left( 1 - \lambda_j(R(\mathbf{A}, \mathbf{B})) \right)^2 + \left( 1 - \lambda_j^{-1}(R(\mathbf{A}, \mathbf{B})) \right)^2
\end{equation}
where $\lambda_j(R(\mathbf{A}, \mathbf{B}))$ is the $j$th largest eigenvalue of the matrix $R(\mathbf{A}, \mathbf{B})$. The function $T$ has valuable properties that may not be immediately obvious.

\begin{proposition}[\citet{ryan2023detecting}]
\label{prop:Tstat prop}
    Let $\mathbf{\Sigma}_1, \mathbf{\Sigma}_2 \in \mathbb{R}^{p \times p}$ be the covariance matrices of data $\mathbf{Z}_1 \in \mathbb{R}^{n_1 \times p}$ and $\mathbf{Z}_2 \in \mathbb{R}^{n_2 \times p}$, respectively, and define $T$ as in (2.5). Then we have that, for any covariance matrix $\mathbf{\Sigma}_0$:

\begin{enumerate}
    \item $T$ is symmetric, that is, $T(\mathbf{\Sigma}_1, \mathbf{\Sigma}_2) = T(\mathbf{\Sigma}_2, \mathbf{\Sigma}_1)$;
    \item $T$ is symmetric with respect to the inversion of matrices, that is,
    \[
    T(\mathbf{\Sigma}_1, \mathbf{\Sigma}_2) = T(\mathbf{\Sigma}_1^{-1}, \mathbf{\Sigma}_2^{-1});
    \]
    \item If $\mathbf{\Sigma}_1 = \mathbf{\Sigma}_0 \mathbf{Z}_1^T \mathbf{Z}_1 \mathbf{\Sigma}_0$ and $\mathbf{\Sigma}_2 = \mathbf{\Sigma}_0 \mathbf{Z}_2^T \mathbf{Z}_2 \mathbf{\Sigma}_0$, then
    \[
    T(\mathbf{\Sigma}_1, \mathbf{\Sigma}_2) = T(\mathbf{Z}_1^T \mathbf{Z}_1, \mathbf{Z}_2^T \mathbf{Z}_2).
    \]
\end{enumerate}
\end{proposition}

The symmetry property is important for a changepoint analysis as the segmentation should be the same regardless of whether the data is read forward or backward. The second property states that $T$ is the same whether we examine the covariance matrix or the precision matrix. This ensures that differences between both small and large eigenvalues can be detected. The third property is particularly important as we can translate \Cref{prop:Tstat prop} from two separate datasets $\mathbf{Z}_1, \mathbf{Z}_2$ to two subsets of a single dataset $\mathbf{X}_{n,p}$. This implies that $T$ provides a test statistic that is independent of the underlying covariance of the data. it is to be noted that the function involves ratio matrices which are widely used in multivariate analysis to compare covariance matrices (\citet{finn1974general}). In particular functions of the eigenvalues of the ratio matrices are standard in literature (\citet{wilks1932certain}, \citet{lawley1938generalization}, \citet{potthoff1964generalized}) for inference and methodologies involving covariance matrices. \Cref{th:th7} also discusses the the LSD of the ratio matrics $R(\mathbf{A},\mathbf{B})$ under suitable conditions. Using the LSD of the ratio matrices, \citet{ryan2023detecting} finds the asymptotic distribution of $T(\mathbf{A},\mathbf{B})$ for two sample covariance matrices as presented in the following theorem, which gives the framework of the changepoint detection method. 

\begin{theorem}[\citet{ryan2023detecting}]
\label{th:th9}
Let \( X_{n_1,p} \in \mathbb{R}^{n_1 \times p} \) and \( X_{n_2,p} \in \mathbb{R}^{n_2 \times p} \) be random matrices with independent not necessarily identically distributed entries \( \{ X_{n_1,i,j}, 1 \leq i \leq n_1, 1 \leq j \leq p \} \) and \( \{ X_{n_2,k,j}, 1 \leq k \leq n_2, 1 \leq j \leq p \} \) with mean 0, variance 1 and fourth moment \( 1 + \kappa \). Furthermore, for any fixed \( \eta > 0 \),
\begin{align}
    \frac{1}{n_1 p} \sum_{j=1}^p \sum_{i=1}^{n_1} \mathbb{E} |X_{n_1,i,j}|^4 \mathbf{1}(|X_{n_1,j,k}| \geq \eta \sqrt{n_1}) &\rightarrow 0 \tag{3.7} \\
    \frac{1}{n_2 p} \sum_{j=1}^p \sum_{i=1}^{n_2} \mathbb{E} |X_{n_2,i,j}|^4 \mathbf{1}(|X_{n_2,j,k}| \geq \eta \sqrt{n_2}) &\rightarrow 0 \tag{3.8}
\end{align}
as \( n_1, n_2, p \) tend to infinity such that $ \frac{p}{n_1} \to \gamma_1 \in (0,1), \frac{p}{n_2} \to \gamma_2 \in (0,1), \gamma = (\gamma_1,\gamma_2)$ and $\mathbf{1}(\cdot)$ denotes the indicator function. Then as \( n \to \infty \),
\begin{equation*}
    T \left( \frac{1}{n_1} X_{n_1,p}^T X_{n_1,p}, \frac{1}{n_2} X_{n_2,p}^T X_{n_2,p} \right) - p \int f^*(x) dF_{\gamma}(x) \rightarrow N(\mu(\gamma), \sigma^2(\gamma))
\end{equation*}
where
\begin{equation}
T(A,B) = \sum_{j=1}^p \big[(1-\lambda_j (B^{-1}A))^2 + (1-\lambda_j^{-1} (B^{-1}A))^2\big] \hspace{0.1cm} (\lambda_j \hspace{0.2cm} \text{is $j$th maximum eigenvalue}),
\end{equation}
\begin{equation}
f^*(x) = (1 - x)^2 + (1 - 1/x)^2,
\end{equation}
\begin{equation}
\mu(\gamma) = 2K_{3,1} \left(1 - \gamma_2 / h^2\right) + 2K_{2,1} \gamma_2 / h + 2K_{3,2} \left(1 - \gamma_1^2 / h^2\right) + 2K_{2,2} \gamma_1 / h,
\end{equation}
\begin{align}
   \sigma^2(\gamma) &= \frac{2(K_{2,1}^2 + K_{3,1}^2 + 2K_{3,2}^2)}{h(h^2 - 1)} + \frac{(J_1 K_{2,1} / h - J_1 K_{3,1} (h^2 + 1))}{h^2 + (h^2 - 1)} \\
&+ \frac{(J_2 K_{2,1}2h) / (h^2 - 1)^3 + J_2 K_{3,1}(1 - 3h^2))}{h(h^2 - 1)^3)} 
\end{align}
\begin{equation}
K_{2,1} = \frac{2h(1 + h^2)}{(1 - \gamma_2)^4 - 2h / (1 - \gamma_2)^2}, \quad K_{2,2} = \frac{2h(1 + h^2)^2}{(1 - \gamma_1)^4} - 2h / (1 - \gamma_1)^2,
\end{equation}
\begin{equation}
K_{3,1} = \frac{h^2}{(1 - \gamma_1)^4}, \quad K_{3,2} = \frac{-2(1 - \gamma_2)^2}{(1 - \gamma_2)^4}, \quad J_2 = (1 - \gamma_2)^4, \quad J_1 = -2 (1-\gamma_2)^2,
\end{equation}
\begin{equation}
h = \sqrt{\gamma_1 + \gamma_2 - \gamma_1 \gamma_2}, \quad \gamma_1 = p / n_1, \quad \gamma_2 = p / n_2,
\end{equation}
\begin{equation}
F_\gamma(dx) = \frac{1 - \gamma_2}{2\pi x (\gamma_1 + \gamma_2 x)} \sqrt{(b - x)(x - a)} \mathbf{I}_{[a,b]}(x) dx,
\end{equation}
\begin{equation}
a = \frac{(1 - h)^2}{(1 - \gamma_2)^2}, \quad b = \frac{(1 + h)^2}{(1 - \gamma_2)^2}.
\end{equation}
\end{theorem}

So, using \Cref{th:th9} we can immediately have a normalized version of $T$, i.e.  
\begin{equation}
\label{eq:normalised Tstat}
    \Tilde{T} = \sigma^{-1}(\gamma) \left(T \left( \frac{1}{n_1} X_{n_1,p}^T X_{n_1,p}, \frac{1}{n_2} X_{n_2,p}^T X_{n_2,p} \right) - p \int f^*(x) dF_{\gamma}(x) - \mu(\gamma) \right)
\end{equation}
which will be asymptotically standard normal, and hence we can use the quantile of standard normal with multiple testing corrections (\citet{haynes2013bonferroni}) to test hypothesis \ref{eq:hypo}. So using \Cref{th:th9}, given one dataset we can test whether the data has one changepoint or not. For the case of Multiple changepoints, the method is generalized using the classic binary segmentation procedure (\citet{scott1974cluster}). 

The binary segmentation method extends a single changepoint test as follows. First, the test is run on the whole data. While running on a particular interval of time $(s,e)$, for each timepoint $\tau$ in that range (except leaving $l$ many timepoints from both sides of the interval, for efficiency purposes as the testing procedure is asymptotic) the algorithm finds the normalized test statistic $\Tilde{T}(\tau)$ (as in \Cref{eq:normalised Tstat}) by breaking the datapoints into two parts pivoting $\tau$ and then finds the maximum value of the test statistic $\Tilde{T}(\tau)$ over $\tau$ in that interval $(s+l,e-l)$ and check if that exceeds a cutoff $\nu$ to guarantee the existence of a changepoint in the interval $(s,e)$. If no change is found then the algorithm terminates. If a changepoint is found, it is added to the list of estimated changepoints, and the binary segmentation procedure is then run on the data to the left and right of the candidate change. This process continues until no more changes are found. Note the threshold, $\nu$, and the minimum segment length, $\ell$, remain the same. Note that several extensions of the traditional binary segmentation procedure have been proposed in recent years (\citet{olshen2004circular}; \citet{fryzlewicz2014wild}) which may be used to generalize the algorithm of \citet{ryan2023detecting}. The full proposed procedure is described in algorithm \ref{algo:RatioBinseg}. 

\begin{algorithm}[H]
\label{algo:RatioBinseg}
\caption{Ratio Binary Segmentation (RatioBinSeg) (\citet{ryan2023detecting})}
\KwIn{Data matrix $X$, interval $(s, e)$, set of changepoints $C$, minimum segment length $\ell$, significance level $\alpha$}
\KwOut{Set of changepoints $C$}

Set $\nu = \Phi^{-1}(1 - \frac{\alpha}{n^2}$), where $\Phi(\cdot)$ N(0,1) CDF\;
\For{$\tau = s + \ell$ \textbf{to} $e - \ell$}{
    Compute $\gamma := \left( \frac{p}{\tau}, \frac{p}{n - \tau} \right)$\;
    Compute $\widetilde{T}(\tau) := \sigma^{-1/2}(\gamma) \left( T\left( \overline{\Sigma}(s, \tau), \overline{\Sigma}(\tau, e) \right) - p \int f^*(x) dF_y - \mu(\gamma) \right)$\;
}
\textbf{end}\\
Set $\hat{\tau} := \arg \max_{\tau} \widetilde{T}(\tau)$ for $s + \ell < \tau < e - \ell$\;
\If{$\widetilde{T}(\hat{\tau}) > \nu$}{
    Set $C_l := \text{RatioBinSeg}(X, (s, \hat{\tau}), C, \ell, \alpha)$\;
    Set $C_r := \text{RatioBinSeg}(X, (\hat{\tau}, e), C, \ell, \alpha)$\;
    Update $C = C \cup \{\hat{\tau}\} \cup C_l \cup C_r$\;
}
\textbf{end}\\
\Return{Set of changepoints $C$}\;
\end{algorithm}

In the algorithm, $\Bar{\Sigma}$ is the natural estimate of $\Sigma$ based on the data in the corresponding time interval.

For the multiple changepoint setting,
let $\tau:= \{\tau_1, . . . , \tau_m\}$ and $\hat{\tau}:= \{\hat{\tau}_1, . . . , \hat{\tau}_m \}$ to denote the
set of true changepoints and the set of estimated changepoints,
respectively. The changepoint $\tau_i$ is said to be detected
correctly if $|\hat{\tau}_j - \tau_i| \le h$ for some $1 \le j \le \hat{m}$ and denote the set
of correctly estimated changes by $\hat{\tau}_c$.
h = 20 is chosen for simulation, although it should be noted that in reality, the desired
accuracy would be application-specific and dependent on the
minimum segment length $l$. Then the False Positive Rate (FPR) is defined as the number of wrongly detected changepoints out of the detected ones, i.e. \[FPR = \frac{|\hat{\tau}| - |\hat{\tau}_c|}{|\hat{\tau}|}\]
Table for FPR for this method and \citet{wang2017optimal} for various $n$ and $p$ are in \Cref{tab:FPR comp}

\begin{table}[h!]
\centering
\caption{Comparison of FPR for various $n,p$} 
\label{tab:FPR comp} 
\begin{tabular}{cccccc}
\hline
\multicolumn{2}{c}{} & \multicolumn{2}{c}{Assumptions of \citet{ryan2023detecting}} & \multicolumn{2}{c}{Assumptions of \citet{wang2017optimal}} \\ \cline{3-6} 
\textbf{p} & \textbf{n} & \textbf{Ratio} & \textbf{Wang} & \textbf{Ratio} & \textbf{Wang} \\ \hline
3 & 500 & 0.24 & 0.63 & \textbf{0.10} & 0.64 \\
3 & 1000 & 0.28 & 0.77 & \textbf{0.14} & 0.80 \\
3 & 2000 & 0.31 & 0.85 & \textbf{0.16} & 0.88 \\
3 & 5000 & 0.31 & 0.90 & \textbf{0.17} & 0.92 \\
10 & 500 & 0.16 & 0.27 & 0.23 & 0.39 \\
10 & 1000 & 0.13 & 0.43 & 0.24 & 0.62 \\
10 & 2000 & 0.10 & 0.53 & 0.24 & 0.75 \\
10 & 5000 & \textbf{0.09} & 0.62 & 0.19 & 0.80 \\
30 & 2000 & \textbf{0.02} & 0.32 & \textbf{0.03} & 0.58 \\
30 & 5000 & \textbf{0.02} & 0.31 & \textbf{0.03} & 0.78 \\
100 & 5000 & \textbf{0.00} & 0.45 & \textbf{0.00} & 0.18 \\ \hline
\end{tabular}
\end{table}

\section{Conclusion and Future directions}
\label{sec: conclusion}

This article highlights the profound role of random matrix theory (RMT) in addressing challenges arising in high-dimensional statistics. By leveraging the asymptotic spectral properties of large random matrices, particularly covariance matrices, and ratios of covariance matrices, RMT provides a novel theoretical foundation for statistical methods. The exploration of both the bulk spectrum and the extreme eigenvalues underscores the versatility of these tools in understanding high-dimensional data structures.

The applications discussed in this article demonstrate the practical relevance of RMT. From inference on covariance matrices to dimensionality reduction through PCA, noise reduction in signal processing, and changepoint detection, RMT proves to be an indispensable framework for tackling modern statistical problems. The unifying principles of RMT not only enhance the theoretical understanding of high-dimensional phenomena but also drive the development of innovative methodologies in diverse fields.

This work provides an inspection of the bridge between the mathematical elegance of random matrix theory and its impactful applications in statistics, emphasizing the potential for further exploration and development in this vibrant intersection of disciplines. We discuss some of the future directions of application of RMT, which has a great potential: 

\begin{itemize}
    \item Most of the results in RMT are based on iid observations. Though work has been done for certain covariance patterns as well, however, there is a great potential for extending the current theory on the eigenvalues of Wishart-type matrices, when the columns of the data matrix can be viewed as a realization of a high-dimensional multivariate time series, and that can have a significant impact on econometrics and finance.
    \item The RMT-based methods can be generalized when there are missing values in the dataset. For example, in high-dimensional spatiotemporal statistics, for each timepoint, the spatial data is in the form of a matrix. In most cases, for each time point, there are a couple of missing values in the matrix consisting of the spatial data at that time point. In literature, in case it is assumed that the spatial data arises from a random field. \citet{deb2017asymptotic} has a detailed discussion about the spectral analysis of such datasets coming from a random field. However, the asymptotic theory provided there assumes the data dimension to be fixed. So generalization of these results using the asymptotic theories of large random matrices is an open problem yet to be solved. 
    \item A potentially useful avenue for the application of RMT is in numerical optimization algorithms that use gradient-based methods for large dimensional data. While there has been explosive growth in mathematical descriptions in the RMT literature, computational tools have not kept pace with the theoretical developments. Integration of computational tools with tools for the analysis of large dimensional data using RMT principles has the potential to create a new paradigm for statistical practices. 
\end{itemize}

\section*{}

\newpage
\section{Appendix}
\textbf{Proof of Theorem 1:} Given $\mathbf{X}_n \sim W_p(\mathbf{\Sigma},n)$ where $n \geqslant p$. Let $\lambda_1^{(n)},\cdots,\lambda_p^{(n)}$ be the eigenvalues of $\mathbf{X}_n$ and $\lambda_1,\cdots,\lambda_p$ be the eigenvalues of $\mathbf{\Sigma}$. For a random variable $X$, let $F_X(\cdot)$ be its CDF and for two CDFs F and G, define 
\begin{equation}
    \Delta(F,G) := \sup_{x \in \mathbb{R}}|F(x) - G(x)|
\end{equation}
We first prove a couple of lemmas needed for the proof. 

\textbf{Lemma 1:} Let $X_n,Y_n,X,Y$ be real-valued random variables having a joint distribution such that $(X_n,X)$ is independent of $(Y_n,Y)$. Then
\begin{equation}
    \Delta(F_{X_n + Y_n}, F_{X+Y}) \leqslant \Delta(F_{X_n},F_X) + \Delta(F_{Y_n},F_Y)
\end{equation}
\textbf{Proof of Lemma 1:} Fix $x \in \mathbb{R}$. Then observe that, 
\begin{flalign*}
     &\qquad|\P(X_n + Y_n \leqslant x) - \P(X + Y \leqslant x)| &&\\
     &\quad= |\E \left(\P(X_n \leqslant x - Y_n) - \P(X \leqslant x - Y) | Y_n, Y \right)| &&\\
     &\quad= |\E (\P(X_n \leqslant x - Y_n) - \P(X_n \leqslant x - Y)  + \P(X_n \leqslant x - Y) - \P(X \leqslant x - Y) \big| Y_n, Y \big)| &&\\
     &\quad= |\E (\P(X_n \leqslant x - Y_n) - \P(X_n \leqslant x - Y)\big| Y_n, Y \big)| + \E\big|\P(X_n \leqslant x - Y) - \P(X \leqslant x - Y) \big| Y_n, Y \big)&&\\
     &\quad= I + II \hspace{0.2cm}(say)&&
\end{flalign*}

Observe that $II \leqslant \Delta(F_{X_n},F_X)$ and 
\begin{flalign*}
    I &= \big|\P(X_n+Y_n \leqslant x) - \P(X_n+Y \leqslant x)\big| \\
    &= \big| \E \big( \P(Y_n \leqslant x - X_n) - \P(Y \leqslant x - X_n)  \big| X_n \big) \big| \\
    &\leqslant \E \big| \P(Y_n \leqslant x - X_n) - \P(Y \leqslant x - X_n) \big| X_n \big)\\
    &\leqslant \Delta(F_{Y_n},F_Y)
\end{flalign*}
Thus for all $x \in \R$, we have $\big|\P(X_n + Y_n \leqslant x) - \P(X + Y \leqslant x)\big| \leqslant \Delta(F_{X_n},F_X) + \Delta(F_{Y_n},F_Y)$. Hence \textbf{Lemma 1} follows. $\blacksquare$ 

\textbf{Lemma 2:} Let $\Phi(\cdot)$ be the N(0,1) CDF and $m$ be a natural number. Then there exists $c >0$ such that
\begin{equation}
    \Delta\left(\Phi \left(\sqrt{\frac{m}{2}} \left( e^{\sqrt{\frac{2}{m}}x} - 1 \right)\right), \Phi(x)\right) \leqslant \frac{c}{\sqrt{m}}
\end{equation}
\textbf{Proof of Lemma 2:} Let $U_1,\cdots,U_m \overset{iid}{\sim} N(0,1)$. Observe that, $Y_m := \sqrt{\frac{m}{2}} \log \left( 1 + \sqrt{2} \cdot \overline{U}_m \right)$ have cdf $F_{Y_m}(x) := \Phi \left(\sqrt{\frac{m}{2}} \left( e^{\sqrt{\frac{2}{m}}x} - 1 \right)\right)$ where $\overline{U}_m := \frac{1}{m} \sum_{i=1}^m U_i$. Consider, $f(x) := \log (1 + \sqrt{2}\cdot x)$. Since $f(0) = 0$ and $f'(0) = \sqrt{2}$, by theorem 2.10 of \citet{pinelis2016optimal}, there exists $c > 0$ such that
\begin{equation}
    \sup_{x \in \R} \left| \P\left( \frac{\sqrt{m}f(\overline{U}_m)}{|f'(0)|} \leqslant x \right) - \Phi(x) \right| \leqslant \frac{c}{\sqrt{m}}
\end{equation}
Since $\frac{\sqrt{m}f(\overline{U}_m)}{|f'(0)|} = Y_m$, we have 
\begin{equation*}
    \Delta(F_{Y_m},\Phi) \leqslant \frac{c}{\sqrt{m}} 
\end{equation*}
 Hence \textbf{Lemma 2} follows. $\blacksquare$

\textbf{Lemma 3:} \hspace{0.2cm}[Berry Esseen type Bounds for log of  $\chi^2$ random variables] Let $Z_m \sim \chi^2_m$. Then, there exists $c > 0$ such that,  
\begin{equation}
    \sup_{x \in \R}\left| \P\left( \sqrt{\frac{m}{2}} \log\left(\frac{Z_m}{m}\right) \leqslant x \right) - \Phi(x) \right| \leqslant \frac{c}{\sqrt{m}}
\end{equation}

\textbf{Proof of Lemma 3:} Observe that by triangle inequality, we have 
\begin{equation}
\label{eq: tineq}
   \Delta(G,\Phi) \leqslant \Delta(G, F_{Y_m}) + \Delta(F_{Y_m},\Phi)
\end{equation}

where $G(x) := \P\left( \sqrt{\frac{m}{2}} \log\left(\frac{Z_m}{m}\right) \leqslant x \right)$ and $F_{Y_m}$ is as in \textbf{Lemma 2}. By \textbf{Lemma 2}, there exists $c_1 > 0$, such that $\Delta(F_{Y_m},\Phi) \leqslant \frac{c_1}{\sqrt{m}}$. By the Berry–Esseen theorem, there exists $c_2 > 0$, such that for all $x \in \R$, 
\begin{equation}
\label{eq: BE phi}
    \Phi(x) - \frac{c_2}{\sqrt{m}} \leqslant \P\left( \sqrt{\frac{m}{2}} \left(\frac{Z_m}{m} - 1\right) \leqslant x \right) \leqslant \Phi(x) + \frac{c_2}{\sqrt{m}}
\end{equation}
Now $G(x) = \P\left( \sqrt{\frac{m}{2}} \log\left(\frac{Z_m}{m}\right) \leqslant x \right) = \P\left(\sqrt{\frac{m}{2}} \left(\frac{Z_m}{m} - 1\right) \leqslant \sqrt{\frac{m}{2}} \left( e^{\sqrt{\frac{2}{m}}x} - 1 \right) \right)$. 

So by (\ref{eq: BE phi}), $\left| G(x) - F_{Y_m}(x) \right| \leqslant \frac{c_2}{\sqrt{m}}$ i.e. $\Delta(G,F_{Y_m}) \leqslant \frac{c_2}{\sqrt{m}}$. Thus by (\ref{eq: BE phi}), $\Delta(G,\Phi) \leqslant \frac{c}{\sqrt{m}}$ where $c:= c_1+c_2$ completing the proof of \textbf{Lemma 3}. $\blacksquare$

Now we complete the proof of the theorem. Observe that $|\mathbf{X}_n| = |\mathbf{\Sigma}|U_1\cdots U_p$ where $U_j \sim \chi^2_{m-p+j}, j=1,\cdots,p$; $U_1,\cdots,U_p$ mutually independent and for a matrix $\mathbf{M}$, $|\mathbf{M}|$ denotes the determinant of $\mathbf{M}$. Thus, 
\begin{flalign*}
    &\qquad \sup_{x\in\R}\left| \P \left( \sqrt{\frac{n}{2p}} \left( \sum_{i=1}^p \text{log}\left( \frac{\lambda_i^{(n)}}{\lambda_i}\right) - \sum_{i=1}^p\text{log} \hspace{0.1cm} (n-p+i) \right) \leqslant x \right) - \Phi(x) \right| &&\\
    &\quad= \sup_{x \in \R} \left| \P \left( \sqrt{\frac{n}{2}}\left( \sum_{i=1}^p \log \left( \frac{U_i}{n-p+i} \right) \right) \leqslant \sqrt{p}x \right) - \Phi(x) \right| &&\\
    &\quad= \sup_{x \in \R} \left| \P \left( \sqrt{\frac{n}{2}}\left( \sum_{i=1}^p \log \left( \frac{U_i}{n-p+i} \right) \right) \leqslant x \right) - \Phi\left(\frac{x}{\sqrt{p}}\right) \right| &&\\  
    &\quad= \sup_{x \in \R} \left| \P \left( \sqrt{\frac{n}{2}}\left( \sum_{i=1}^p \log \left( \frac{U_i}{n-p+i} \right) \right) \leqslant x \right) - \P(Y_1+\cdots+Y_p \leqslant x) \right| \hspace{0.2cm} ,Y_1,\cdots,Y_p \overset{iid}{\sim} N(0,1) &&\\
    &\quad \leqslant \sum_{i=1}^p \sup_{x\in \R}\left|  \P \left( \sqrt{\frac{n}{2}}\left( \sum_{i=1}^p \log \left( \frac{U_i}{n-p+i} \right) \right) \leqslant x \right) - \Phi(x) \right| \hspace{0.2cm} (\text{by \textbf{Lemma 1}}) &&\\
    &\quad \leqslant C \sum_{i=1}^p \frac{1}{\sqrt{n-p+i}} \hspace{0.2cm} (\text{by \textbf{Lemma 3}}) &&\\
    &\quad = O\left(\frac{p}{\sqrt{n}}\right)
\end{flalign*}

Hence \textbf{Theorem 1} follows. $\blacksquare$
\end{document}